# THE FOUNDING OF ARCETRI OBSERVATORY IN FLORENCE

Simone Bianchi

INAF-Osservatorio Astrofisico di Arcetri,
Largo E. Fermi, 5, 50125, Florence, Italy
simone.bianchi@inaf.it

**Abstract:** The first idea of establishing a public astronomical observatory in Florence, Capital of the Grand Duchy of Tuscany, dates to the mid of the 18th century. Initially, the use of a low building on a high ground was proposed, and the hill of Arcetri suggested as a proper location. At the end of the century, the Florence Observatory - or *Specola* - was built instead on a tower at the same level as the city's centre. As soon as astronomers started to use this observatory, they recognized all its flaws and struggled to search for a better location.

Giovanni Battista Donati, director of the *Specola* of Florence from the eve of the Italian Unification in 1859, finally succeeded in creating a new observatory: first, he obtained funds from the Parliament of the Kingdom of Italy to build an equatorial mount for the Amici 28-cm refractor, which could not be installed conveniently in the tower of the *Specola*; then, he went through the process of selecting a proper site, seeking funds and finally building Arcetri Observatory. Although Donati was a pioneer of spectroscopy and astrophysics, his intent was to establish a modern observatory for classical astronomy, as the Italian peninsula did not have a national observatory like those located in many foreign capitals – Florence was the capital of the Kingdom of Italy from 1865 to 1871. To promote the project, Donati made use of writings by one of the most authoritative European astronomers, Otto Wilhelm Struve.

The paper describes all these steps, eventually leading to the final inauguration of the Arcetri Observatory in 1872, almost 150 years ago.

## 1 ESTABLISHING AN ASTRONOMICAL OBSERVATORY IN FLORENCE

It was only in the eighteenth century that public astronomical observatories were established in the Italian peninsula, which at the time comprised several independent states. The first observatory was that of the Institute of Science of Bologna, in the Papal States, whose observing tower was completed in 1725 (Zanini, 2009). The second Italian state to have a public observatory was the Grand Duchy of Tuscany, ruled by Gian Gastone (1671-1737) of the House of Medici. The observatory (or *specola* as these institutions were originally called in Italian) was built in Pisa between 1735 and 1746; it belonged to the local University (Di Bono 1990).

### 1.1 Florence vs Pisa

After the death of Gian Gastone, the Medici line extinguished and Francis Stephen (1708-1765) of the House of Lorraine became the Grand Duke as Francis II. He lived in Vienna, where he had married Maria Theresa of Habsburg, and in 1745 he became Holy Roman Emperor, in co-regency with his wife. During his reign, Francis II left the administration of Tuscany in the hands of a Regency Council, which sat in the Capital of the Grand Duchy, Florence. The Council asked the opinion of Tommaso Perelli (1704-1783), professor of astronomy in Pisa and director of the Observatory from 1739, about establishing a new astronomical observatory in Florence, on the upper floor of the building hosting the church of Orsanmichele (Figures 1 and 2). Perelli complied with a report dated 20 February 1751. The astronomer judged the stability and space inside the building sufficient, even though its position in the city centre limited the sky view, because of high towers such as that of Palazzo Vecchio and of the domes of the cathedral of St. Maria del Fiore and of the church of St. Lorenzo (Figures 1 and 2). But instead, Perelli suggested:

> … let us use one of the very pleasant suburban hillocks of which Florence is surrounded, following in this the example of the English who established their Observatory on a hill named Greenwich a mile away from London ... The multitude of villas scattered throughout the Florentine countryside makes the purchase of a proportionate and conveniently located building very easy and will involve little expenditure … every mediocre building is sufficient, the same ground rooms, and any terrace, even a little above the ground, being able to serve to install astronomical instruments … if the choice were left to me, I would be inclined to select the Arcetri hill, a place ennobled by the observations of the great Galileo and the many years that he stayed there … (Corsini 1924:264)[1].

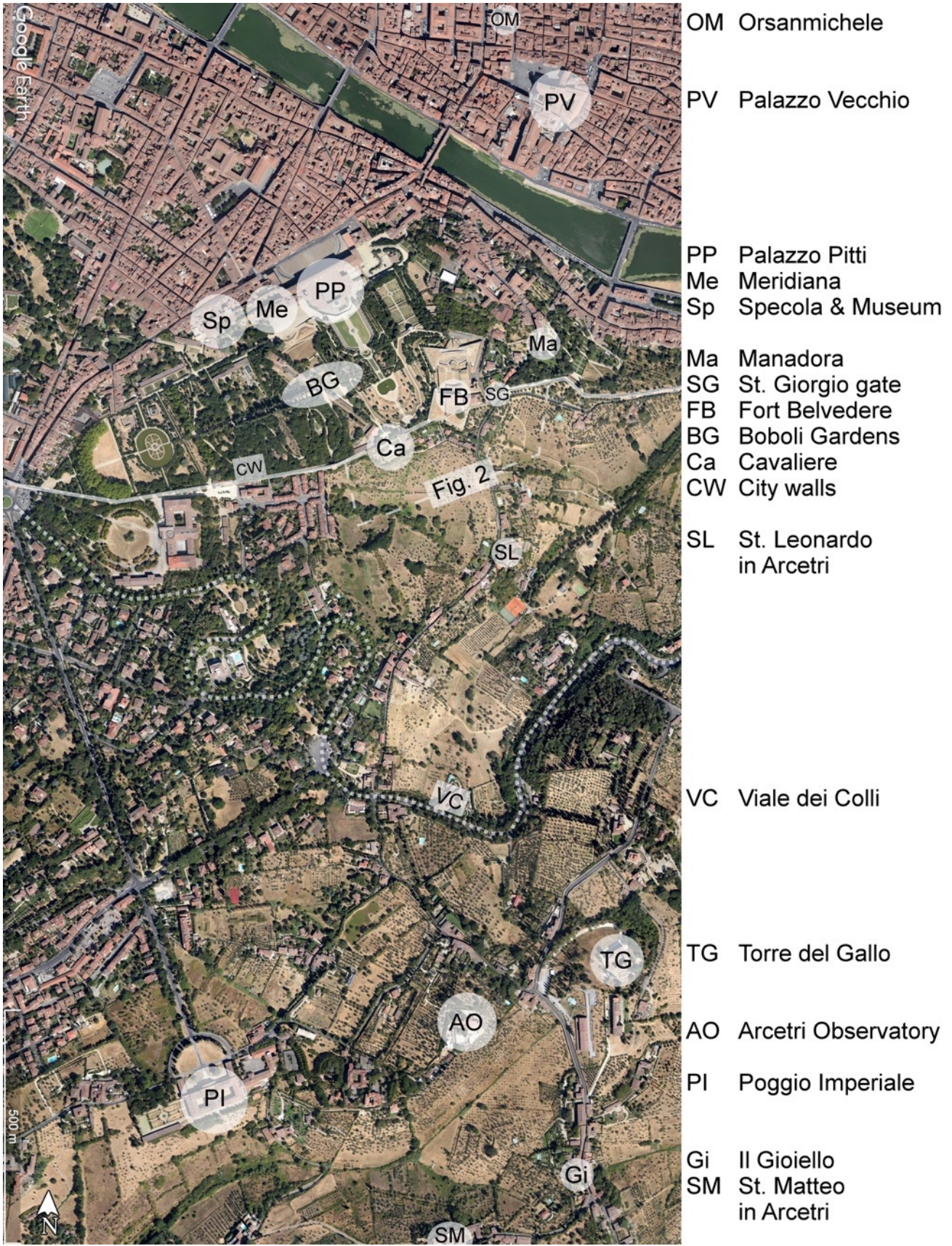

**Figure 1:** Aerial view of Florence from the city centre in the North to the Arcetri area in the South (Google Earth).

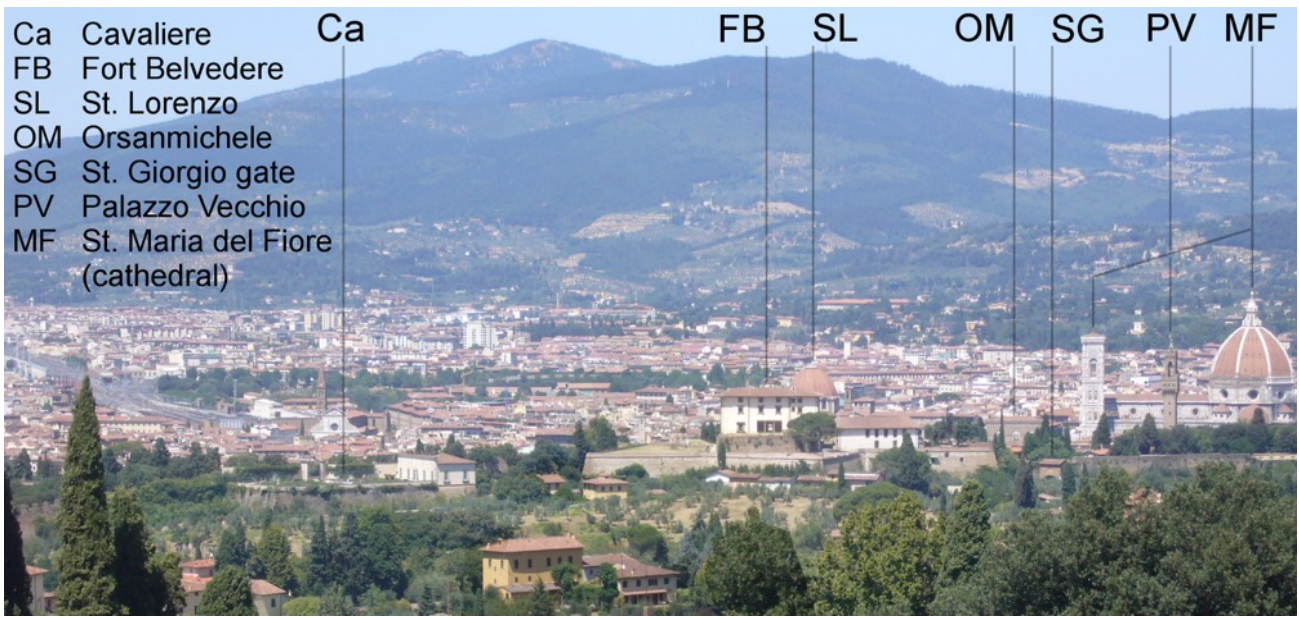
**Figure 2:** A view looking north from the roof of Arcetri Observatory (the extent of the view is shown in Figure 1).

The name Arcetri refers to a small area just south of Florence's St. Giorgio hill gate. Two churches are named "in Arcetri", those of St. Leonardo and of St. Matteo. The latter was part of the convent in which the two daughters of Galileo Galilei were secluded, and is close to the villa Il Gioiello (Figure 1), where the astronomer lived from 1631, was confined by the Inquisition from the end of 1633, and eventually died in 1642 (Godoli et al. 2017). Thus, Arcetri was suitable for astronomy not only because of the unimpeded view of the horizon from the hilltop, but it was also an inspirational location.

Perelli concluded his report by discussing the advantages of practicing astronomy in Florence rather than in Pisa. Beside the air quality (Pisa was often foggy), the Capital had the advantage of a larger population: not only were there more skilled artisans in Florence for the maintenance and repair of instruments, but there were more persons interested in astronomy and in conducting observations. Instead, Pisa was a small city where the large number of students attending the University only resided during term-time, and they were too busy with their studies to have time for astronomy. Furthermore, most of the professors were either old or not learned in the subject, and the few that were did not collaborate because of academic jealousy. Finally, Perelli suggested following the example of the "most learned Nations", which had established Observatories in their main cities and capitals, such as London, Paris, Vienna, Copenhagen, Saint-Petersburg and Beijing, or, without going too far afield, Bologna, the largest city in the Papal States besides Rome.

The disaffection of Perelli towards his own Observatory and the academic environment in Pisa is evident in his comments. Perelli was reputedly a learned man in many fields, but was negligent towards his observing duties and academic obligations (Di Bono 1990). In fact, it has been argued that he suggested himself the idea of a new observatory in Florence, upon which he was asked to report (Corsini 1924). He continued to support this project in the following years. Around the spring of 1760, he proposed the Regency Council to use as a *specola* a villa close to the gate of St. Giorgio, called Manadora (now Villa Bardini; see Figure 1), which had been temporarily confiscated by the government. He offered to pay for an annual rent and provide a few "excellent astronomical instruments" he had bought from England with his own money, which would better be used in Florence to provide astronomical education for the "noble Tuscan youth" than in the insalubrious and foggy air surrounding the Pisa Observatory (Corsini 1924:260-261; Perelli 1760).

About a decade later, Perelli repeated his arguments in favour of a Florentine Observatory in what one would deem a very inappropriate place: a text intended as a preface to the first volume of observations from Pisa (Perelli 1769); not unsurprisingly, the volume was printed with a preface written by Giuseppe Antonio Slop (1740-1808), assistant of Perelli (Di Bono 1990). In his text, Perelli said that it would have been better to build an observatory on an old tower on the hill close to the church of St. Miniato in Florence. This description seems to relate to a medieval building named Torre del Gallo, located on the highest point of the Arcetri hill (Figure 1; St. Miniato, which is not in this figure, is about 400 m from the middle of its eastern border and 1000 m NNE of the Torre del Gallo). An historical account of the Grand Duchy under the House of Lorraine says that Perelli "wanted, since the times of the Regency, to erect a similar observatory on the tower of Or-San-Michele, or on that named *del Gallo*" (Zobi 1850:331).

**1.2 The *Specola* of the *Imperiale & Reale Museo di Fisica e Storia Naturale***

In 1765 Peter Leopold (1747-1792) became Grand Duke of Tuscany, a post held until 1790 when he was crowned Holy Roman Emperor. In contrast to his father, he directly ruled the country from Florence, and promoted several state reforms. In 1766 he assigned the physicist Felice Fontana (1730-1805) the task of gathering together all of the scientific instruments and specimens scattered throughout the Royal Collections. This activity led to the establishment of the Imperial and Royal Museum of Physics and Natural Sciences, of which Fontana became Director. The Museum opened to the public in 1775, in a building next to the monarch's residence, the Palazzo Pitti (Figure 1).

The Museum included laboratories and an astronomical observatory (Figure 3), for which a new octagonal tower was erected between 1780 and 1789 (Schiff 1928). The naturalist Giovanni Fabbroni (1752-1822), Deputy-Director of the Museum, had suggested installing the observatory on a low building in the Cavaliere, a rectangular garden located on top of a bastion on the city walls. The Cavaliere is on the higher part of the Boboli Gardens, which lie on a slope rising from Palazzo Pitti and the Museum up to the Fort Belvedere (Figure 1). However, Fontana preferred the tower solution, because of contiguity with the Museum (Miniati 1984). In fact, there is no mention in the Museum projects of Perelli's suggestions to use a higher site. As a result, the sky view from the new *Specola* was limited: "its horizon [is] dominated, with the exception of one side, by buildings and nearby hills" (De' Vecchi 1808: 7), and with "The Fort Belvedere, the decoration and the trees of the Boboli Gardens hide a considerable, if not at least an important, portion of this [eastern] part of the sky" (De' Vecchi 1810: 29).

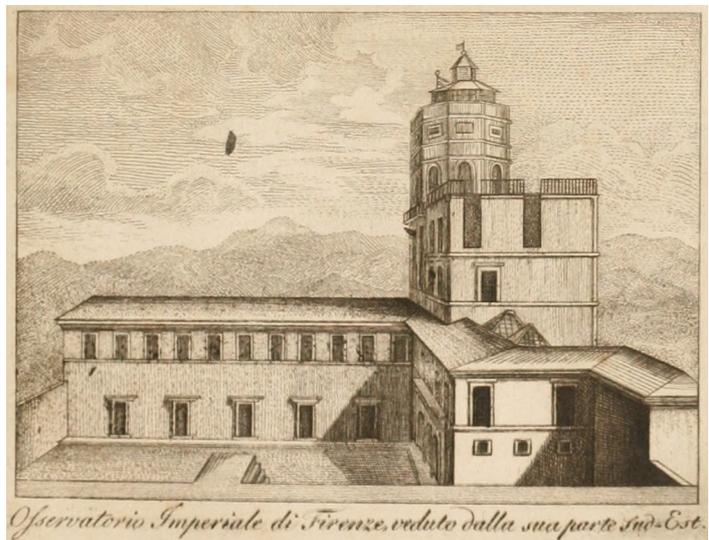

**Figure 3:** "The Imperial Observatory of Florence, seen from its South East side" (Annali 1810: frontispiece).

At the time, the Italian peninsula lacked skilled manufactures of scientific instruments, so Fontana and Fabbroni spent from 1775 to 1780 travelling abroad and acquiring the equipment for the Museum (Schiff 1928). Most of the astronomical instruments were bought in London, where the two Italians arrived at the beginning of 1778. Among the main instruments that the Museum got from London were: a transit instrument and a zenith sector by Jeremiah Sisson (1720-1783), both of about 10-cm aperture (4"); an 8-cm (3") achromatic telescope by Peter Dollond (1731-1820); a 15-cm (6") reflecting telescope by William Herschel (1738-1822); and a pendulum by Larcum Kendall (1719-1790) (De' Vecchi 1810; Miniati 1984). In London Fontana also met Jesse Ramsden (1735-1800) and they discussed the types of instruments that could make the new observatory a world leader. Among these we find an astronomical circle of 3.6-m (12-feet) diameter, a giant instrument that Ramsden had never made (and would never make; see McConnel 2007). Back in Tuscany, Fontana dreamt of constructing the circle himself, in order to have "the first, the greatest, the most useful instrument made in Europe, and perhaps the only one for a few centuries" (Miniati 1984: 213). A wooden model of the instruments was made in 1796, following the design of the Ramsden circle of the Palermo Observatory. However, the instrument was never realized, due to the lack of technical skill, financial problems and last, but not least, personal rivalries between Fontana and Fabbroni (Miniati 1984).

Another problem for Florence Observatory was the lack of a resident astronomer. No astronomer supervised the planning and construction of the building, with the exception of Johann III Bernoulli (1744-1807), royal astronomer in Berlin, who proposed a few modifications to the original project when visiting Florence in 1775. Slop, who had become director of the Pisa Observatory, gave assistance in 1783-1784 with the installation of the Sisson instruments and in determining a meridian line (De' Vecchi 1810). However, for a long time thereafter the instruments were not used or maintained.

The first director of Florence Observatory and professor of astronomy, Domenico De' Vecchi (1768-1852), was appointed only in 1807, when Maria Luisa of Spain (1782-1824), Regent of the Napoleonic Kingdom of Etruria, opened the Museum to teaching. De' Vecchi could not but note the limits of the observatory concerning the visibility of the sky, the stability of the building, and its odd layout. After verifying the state of the instrumentation, he started the first program of observations: the geographical coordinates of the Observatory were measured and a preliminary stellar catalog was begun. Yet, De' Vecchi's activity did not last long. In 1814 the Grand Duchy was restored, and in a frenzy to undo all of the acts of the French administration teaching in the Museum was ended and all its professors were dismissed (Schiff 1929). Afterwards, the *Specola* was entrusted to the astronomers of a private observatory in Florence, which had been established by Leonardo Ximenes S. J. (1716-1786) in 1756 and bequeathed to the Piarist school (Bravieri 1985). Those astronomers seldom used the *Specola*. Nevertheless, new instruments were bought during this period: a refractor with a 11-cm (4") Fraunhofer doublet and a repeating circle by Reichenbach (Funaro 2001; Inghirami 1819).

Hopes for a change occurred when Leopold II (1797-1870) ascended to the throne of Tuscany in 1824. Leopold II had a genuine interest for science, reflecting that of his grandfather Peter Leopold (Funaro 2001). A year later he appointed as director of the *Specola* Jean Louis Pons (1761-1831). Pons had started his career as a comet hunter in Marseille, then in 1819 he was called by Maria Luisa of Spain to direct the new Observatory of Marlia in the Duchy of Lucca (Arrighi 1956), the new state the Restoration had assigned to her. After the death of Maria Luisa, the Marlia Observatory was closed and Pons moved to Florence, where he found seven more comets, among the 37 discoveries he made during his lifetime (Kronk 2003). Pons' feats finally drew the attention of the international astronomical community to Florence, which became the "quartier général des comètes" (von Zach 1825: 187).

The fate of the ephemeral Observatory of Marlia was soon followed by that of Pisa, which was dismantled in 1831 (the last observations having been carried out in 1807; see Di Bono 1990). The *Specola* of Florence remained the only extant public observatory in the Grand Duchy (the Duchy of Lucca was united with the rest of Tuscany in 1847), but by that time, the Italian peninsula counted almost a dozen such institutions (Pigatto 2012).

**2 DREAMING ABOUT A NEW OBSERVATORY**

While Pons' work in astronomy did not need special requirements, the Museum authorities had nevertheless the desire to improve the conditions at the Observatory. This is evident in the actions of the physicist Vincenzo Antinori (1792-1865), who was Director of the Museum from 1829 (Figure 4).

**2.1 A first project for the Belvedere**

In September 1830, Antinori and Giovanni Battista Amici (1789-1863) visited Fort Belvedere (Meschiari 2005). Amici (Figure 5) was the only instrument-maker in the Italian peninsula who could rival foreign manufacturers. He was well known in Europe for his microscopes and telescopes, and in particular, his Newtonian reflectors had been praised for their quality by John Herschel (1792-1871; e.g. see Herschel 1825). In 1827 he furnished the Observatory of his hometown, Modena, with a complete set of astronomical instruments (Meschiari & Bianchi 2013). During his visit to Florence, Amici had promised to send Antinori "the description and plans necessary for the good construction of a modern astronomical observatory" (Antinori 1830), and he complied by sending:

> … the description of the Fraunhofer telescope published by Struve from which you can recognize the layout of the rotating roof that covers the room that houses that magnificent instrument. In dealing with a new observatory, it will not be necessary to be slavish imitators of the constructions of the others, but it is good to examine what has been made by others in order to choose the best (Amici 1831a).

Clearly, the aim was to establish a modern observatory similar to that at Dorpat (presently Tartu in Estonia) where Wilhelm Struve (1793-1864) had installed a 24-cm (9-inch) refractor by Fraunhofer on an equatorial mounting, the best such instrument made up to that date (Struve 1825), and possibly to make it without relying on foreign workshops. Soon Antinori replied by saying that the Grand Duke seemed to be in favor of such a project, and had already ordered an investigation of Fort Belvedere in order to locate solid bedrock beneath the soil. Antinori continued by providing a preliminary sketch of a rectangular building, with two domes (one on each end), to be built on the bastions of the Fort (Antinori 1831a).

More details on the project for Fort Belvedere can be found in a proposal sent by Antinori to the Grand Duke on 15 February 1832, soon after Amici had succeeded Pons to the direction of the *Specola* at the end of 1831. Antinori reminded Leopold II of all the defects of the *Specola* building, from the limited view to the unstable meridian hall, the limited space and the lack of a house for the astronomer. Some of these defects could be mitigated, but only at the cost of significant modifications to the tower. Instead, Antinori suggested that they abandon the *Specola* and install a new observatory on the sixteenth-century central building of the Fort. On the roof of that building, which was solidly built and well oriented across the meridian line, there would be space for two domes, one for an equatorial instrument (which the *Specola* did not have yet, unless Antinori considered a portable, 5-cm (2") aperture, instrument by Dollond; De Vecchi 1810) and one for the repeating circle, while between the two domes one could build a gallery to host meridian and other instruments. One of the floors below the observation deck could be used for the library, study and workshop of the observatory, and another for the astronomer's house. Antinori concluded the letter to his Monarch with the following flattering description:

> … the new observatory could be one of the most magnificent in Italy given the location, the vastness horizon, the stability of the building, the excellence of the instruments, the commodities it could have, and His Highness will then have the glory of raising a monument in the homeland of the great Galileo worthy of the restorer of modern astronomy, and of assigning to the education of his subjects a building [the Fort] that was raised by the Medici with quite another intention (Antinori 1832).

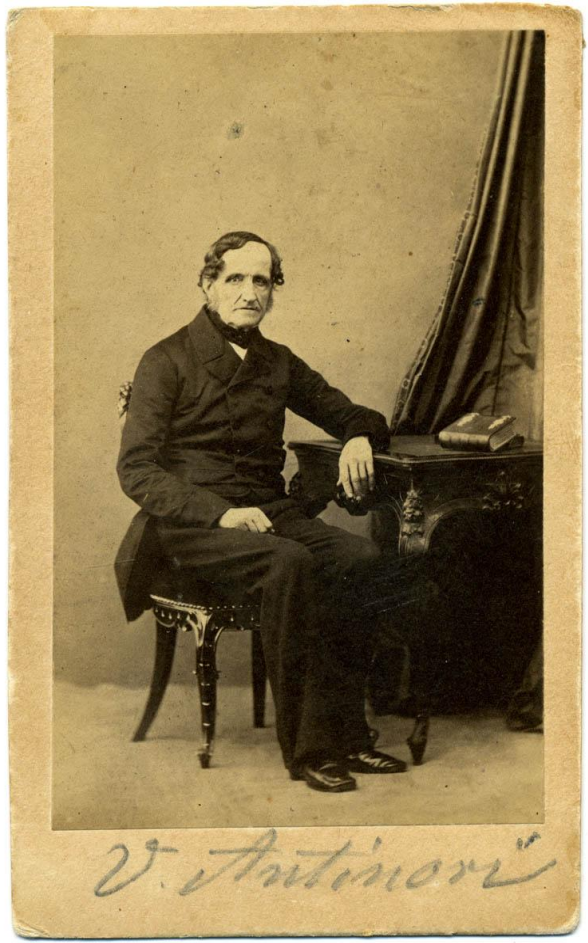

**Figure 4**: Vincenzo Antinori (ca. 1860; Museo Galileo, Florence; Raccolta fotografica Cartes-de-visite raffiguranti medici e scienziati).

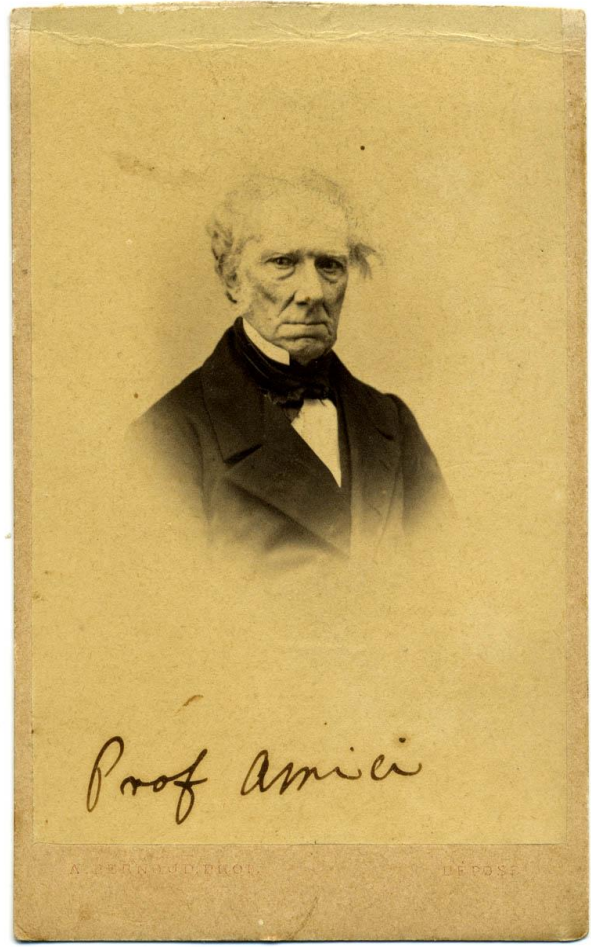

**Figure 5**: Giovanni Battista Amici (ca. 1860; Museo Galileo, Florence; Raccolta fotografica Cartes-de-visite raffiguranti medici e scienziati).

### 2.2 An instrument-maker as Director

When Amici moved to Florence at the end of 1831, he moved also his private workshop that he then installed in his house in the city (Meschiari 2005). He also used the Observatory as a showcase for the instruments he produced. For example, when the Director of Vienna Observatory, Karl Ludwig von Littrow (1811-1877) visited the *Specola* he found many instruments on display, several of which were the private property of Amici (Bianchi 2010). Amici practiced astronomy mainly as a way to test the products of his workshop, and in Florence he did not use them to attempt to carry out any systematic program of observations. This drew the criticism of some contemporaries: "he attended little his post, never taught science, seldom came to the Museum; more than anything else, he attended to the workshop he had in his house …" (Parlatore 1992: 270); "although he was a very respectable man, Amici did not have the intention of being an astronomer" (Donati 1863b).

However, it had been clear from the beginning that Amici was not called to Florence as an astronomer, but as an instrument maker, and that he could continue his own private enterprise. This is evident from the letter in which Antinori announced that Leopold II called Amici to the post of director. The offer gave ample freedom to Amici: he could move his instrument workshop to Florence and continue to attend to it; given his poor health condition, he could for the moment "supervise the construction of the building of the new observatory, which should soon begin, and build the instruments that will be needed for the equipment of the new Specola", leaving perhaps observations to one of his sons (Antinori 1831b). Amici accepted these conditions, and admitted: "I indeed delighted myself in some branches of Astronomy, but I'm no astronomer" (Amici 1831b). In the project for the observatory in the Fort, Antinori (1832) asked that space be allocate "… for his [Amici's] optical and mathematical workshop, as it will be too uncomfortable for him to descend to the city to supervise and direct the works…".

Notwithstanding these grand plans, the project for Fort Belvedere did not proceed. In a report on the status of the observatory and its activity (or better, inactivity) sent by Antinori to the Grand Duke a decade later, there is no mention of a new location (Antinori 1840). The instrumentation had improved little: there was now an equatorial instrument, obtained by mounting together pieces of old instrumentations with a 10-cm telescope by Amici (Bianchi 2010) – but it had still to be installed. Meanwhile, funding for a meridian circle was requested, so that measurements could be made more quickly than with the repeating circle. A suitable research program was defined (comet searches, observations of lunar occultations, sunspots, shooting stars, double stars, stellar proper motions and – possibly – parallaxes), which could be done "… with profit and without expenses …" by young men educated in mathematics and willing to learn celestial mechanics. Perhaps Antinori was referring to students in their last years of academic studies, since the Museum had reopened for lectures in 1833.

### 2.3 The Amici telescope

The only major request described in Antinori's (1840) report was that a movable roof be built over part of the greenhouse in the botanical garden, so that this could house a big telescope, built around the achromatic objective that the Grand Duke had just funded. The 28-cm (11-inch) doublet objective, made from crown and flint blanks from the Guinand factory in Paris, was polished in the workshop of the Museum under Amici's supervision (Meschiari 2006; Righini 1969). A first version was presented at the 3$^{rd}$ Meeting of the Italian Scientists, held in Florence in 1841, but the crown glass was defective and a new lens was polished later. With this delay, the telescope, mounted on a mahogany tube, was only completed in 1845 (Antinori 1849). Initially the telescope was mounted with a tripod on the eyepiece side and a wooden pillar on the objective side, but with its large dimension – the objective has a focal length of 5.3 metres – it was difficult to use in the observing room in the tower of the Specola (Figure 6), so it was moved to the adjoining terrace by sliding it on a metal sheet placed under its mounting (Bianchi 2010). Despite its limitations, the Amici telescope was the largest refractor available in Italy, and it would remain as such for almost 40 years.

A continuous program of astronomical observations started at the Specola only in August 1852, when the young Giovanni Battista Donati (1826-1873) was appointed apprentice astronomer (Figure 7). Donati started using the Amici telescope for comet and asteroid observations in 1854 – the first documented use of the instrument (Meschiari & Bianchi 2013). However, using the telescope was not easy: when describing observations of comet 5D/Brorsen in 1857, Donati (1857: 357) stated that the telescope "… would be very useful for astronomical observations if one did not have to contend with the difficulty of moving it properly, because of its provisional mounting". One might wonder if the publishing of such a comment in a scientific journal was meant to draw attention to the shortcomings of the Observatory. In fact, at the end of 1858 Antinori proposed to erect a rotating dome on top of a new Museum annex that was being built to host a recently acquired botanical collection (and the necessity of this dome was also mentioned in Antinori 1849). Without this dome, which could not fit easily to the top of the tower of the *Specola*, it was not possible to have an equatorial mounting with a clock drive for the Amici telescope, which was "one of the best and most powerful in existence; but which … does not give to science the service that it should and largely remains wasted" (Antinori 1858).

### 3 PLANNING THE NEW OBSERVATORY

On 27 April 1859 Leopold II left Florence, having refused to support the Italian independence cause and to join the 2$^{nd}$ Italian War of Independence on the side of Piedmont against Austria. A series of provisional governments then ruled Tuscany through the 1860 plebiscite for the annexation to Piedmont and other Italian provinces, and the proclamation in 1861 of the Kingdom of Italy under Victor Emmanuel II (1820-1878) of the House of Savoy.

In December 1859, the Tuscan government founded in Florence an advanced academy, the Institute of Superior Studies, and the Museum became the section of Natural Sciences of this Institute. The Minister of Public Instruction of the Tuscan government was the agronomist Cosimo Ridolfi (1794-1865), a former student of the Museum (Figure 8). A few months later, Ridolfi succeeded to Antinori, who had resigned from the Museum Directorship, probably as a sign of loyalty to Leopold II. At the start of the Institute, Amici also was removed from his post and Donati was promoted Professor of Astronomy and Director of the Observatory.

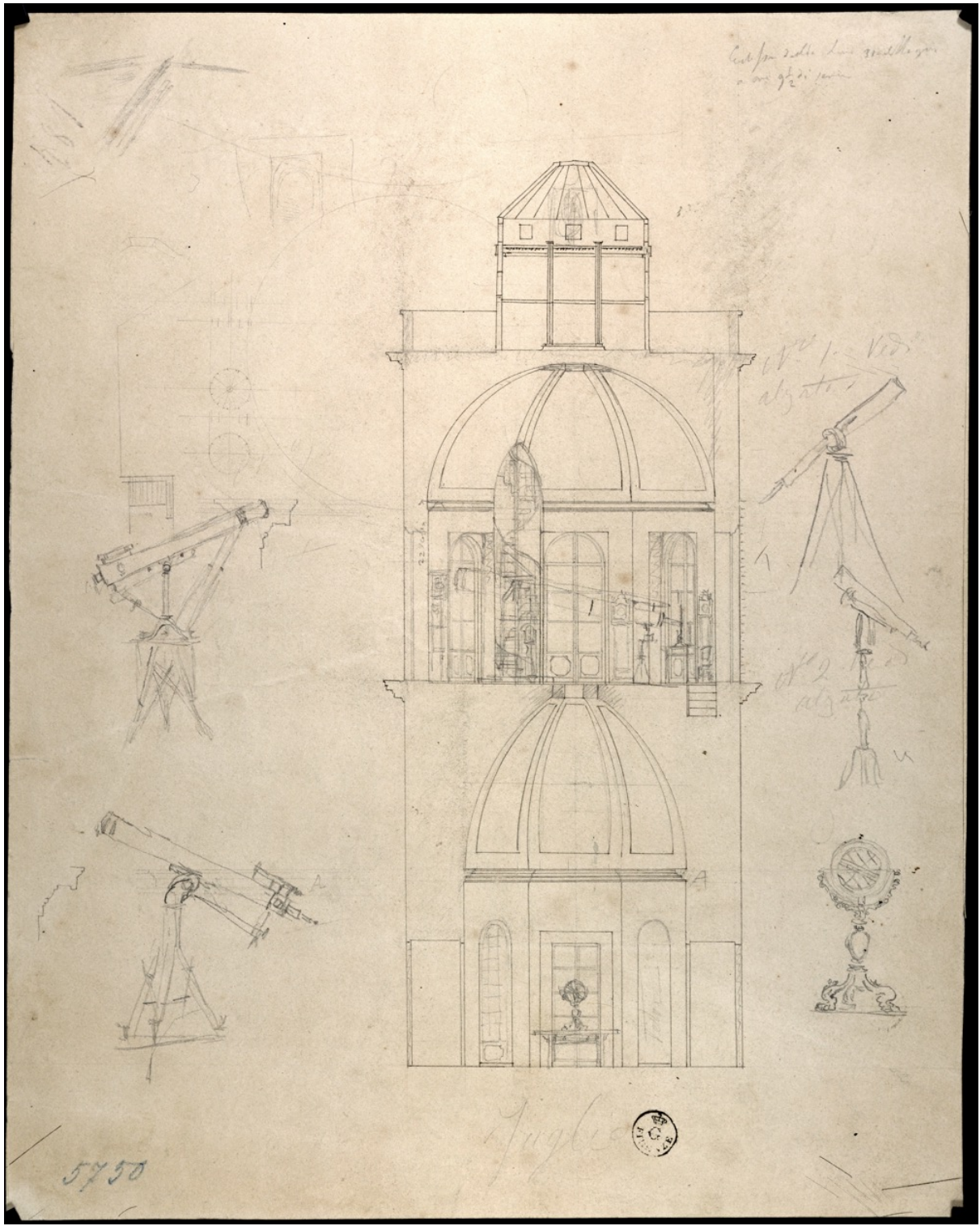

**Figure 6:** Vertical section of the Tower of the *Specola* in Florence (Giuseppe Martelli 1792-1876. Gabinetto Disegni e Stampe Uffizi, Firenze, inv. 5750A). A note on the drawing refers to the Lunar eclipse of 31 May 1844. Over the drawing of the upper observing room is a sketch of a large refractor, likely the Amici telescope (or its project).

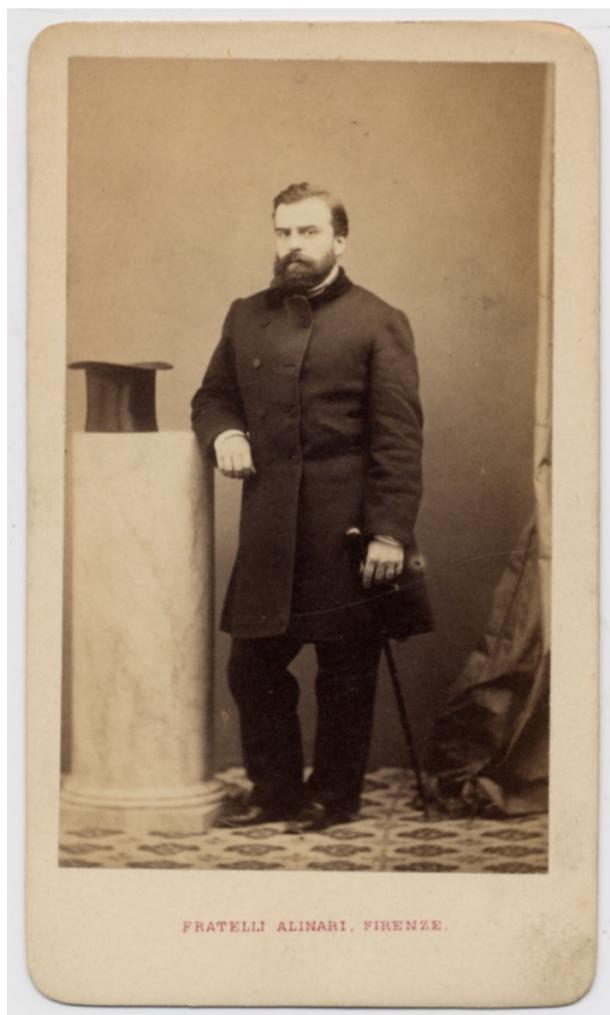 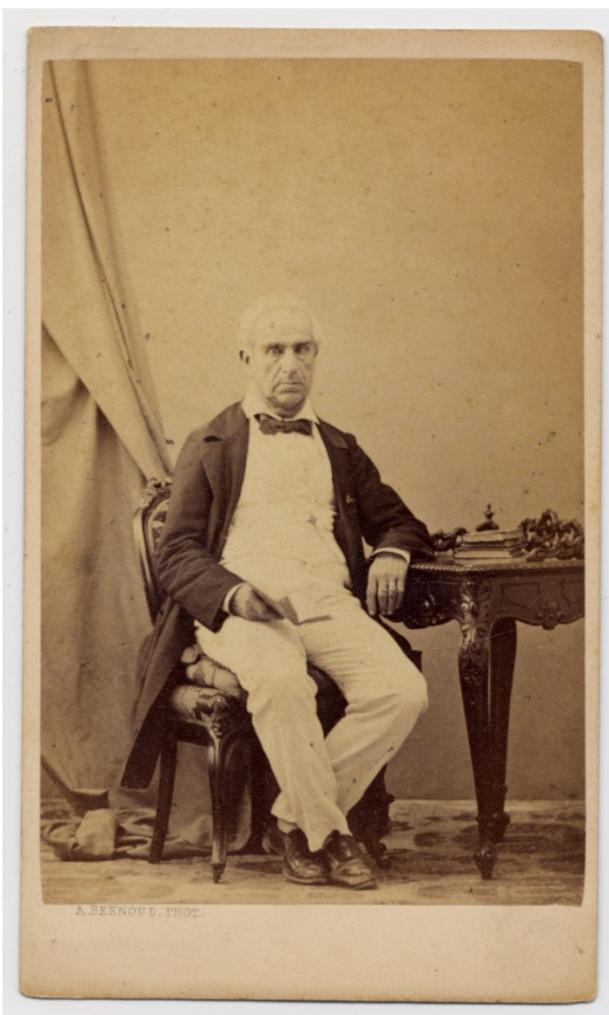

**Figure 7**: Giovanni Battista Donati (F.lli Alinari, ca. 1860; Courtesy: Alinari Archives. Biblioteca Malatestiana, Cesena; Fondo Comandini, FFC 229).

**Figure 8**: Cosimo Ridolfi (A. Bernoud, 1865; Biblioteca Malatestiana, Cesena; Fondo Comandini, FFC 596).

Donati had been adjunct astronomer since 1858, when he became worldwide famous for the discovery of C/1858 L1, one of the most beautiful comets of the century (Gasperini et al. 2011). His initial interest had been to search for new comets - he discovered five - and the calculation of their orbits (Galli et al. 2013). At the turn of the decade, he observed the spectra of stars of different colours and became a pioneer of astrophysics, providing the first crude classification of stellar spectra (Bianchi et al. 2016; Donati 1862; Donati 1863a). Donati and Ridolfi soon started to collaborate to carry out the projects that Amici and Antinori had not been able to achieve.

**3.1 The conditions of the Italian Observatories**

Besides the limited sky visibility, the lack of space and the poor stability of the tower of the *Specola*, its position within the city made it vulnerable to the results of human activity such as smoke, vibrations due to vehicles, and light pollution (Cipolletti 1872; Donati 1866a). The same problems were shared by almost all the other observatories in the Italian Peninsula which, with the notable exception of the Capodimonte Observatory in Naples, were located on high buildings within city centres. After the Unification, the Kingdom of Italy thus found itself with many unsuitable observatories (for an account of the state of Italian astronomy in those years see Bònoli et al. 2005). This problem was not limited to astronomy, but extended to the plethora of institutes (such as libraries and universities) inherited from the many pre-unitarian states. Donati was soon involved in discussions on how to rationalise the funding provided for astronomy (Bianchi & Galli 2014). He was in favour of establishing a single National Observatory (possibly, the new observatory for Florence that he was planning to build). Yet, competition and jealousy among Italy's professional astronomers meant that the other existing Italian observatories could not be simply ignored (Pigatto 2012).

In 1862, Donati took part in a meeting in the capital of the Kingdom, Turin, organised by the Minister of Public Instruction, the physicist Carlo Matteucci (1811-1868), who had been his professor at Pisa University. Matteucci also invited, among others, the directors of the Observatory of Padua, Giovanni Santini (1787-1877), and of the Collegio Romano in Rome, Angelo Secchi S. J. (1818-1878), that were not yet part of the Kingdom. The participants agreed that the ideal solution - a single, new, observatory - would be too costly for the limited finances of the new state (Bianchi & Galli 2014), and a more practical solution would be to concentrate most of the investment on four observatories, evenly distributed throughout the peninsula. This is how Donati summarised the conclusions of the meeting:

> In 1862 ... it was established that in Italy the Observatories that should be maintained and expanded were only those of Milan, Florence, Naples and Palermo. ... No country has more Observatories than Italy and nevertheless no country is more unfortunate than ours because none of the current Italian Observatories is equipped with instruments and located in such a way as to be able to be on par with the Observatories of other Nations. We should do to our Observatories what we have done to our past governments, tear them all down to build a truly National one, that addresses the needs of the times that have changed both for politics and for science. Perhaps, however, a single national Observatory for Italy would be too little, given the favorable conditions of our geographical position and of our climate; but keeping four of them is already more than enough if we wish to cultivate the very important science of Astronomy with real profit (Donati 1864a).

The idea of restricting the number of first-class observatories to four was proposed again a decade later, but the reformation project never proceeded (Poppi et al. 2005). Italy maintained the peculiarity of having many independent observatories until the formation of the National Institute for Astrophysics in 1999 (Chinnici 2015; Pigatto 2012).

**3.2 The scope of the new Observatory**

The astronomers who met in Turin in 1862 also highlighted the need for a meridian circle installed in a stable, ground-level building, something that was not yet available in Italy. The need for modern instrumentation that could be used for high-precision measurements of stellar positions also emerged from the geodetic work that the Italian astronomers were to carry out within the program of the *Mitteleuropäische Gradmessung* enterprise. At the first general conference of the association, held in Berlin in October 1864, hope was expressed that observatories in Italy and Switzerland could extend fundamental measurements to stars at southern declinations (Förster 1865). Yet the members of the Italian commission, including Donati and Giovanni Virginio Schiaparelli (1835-1910), director of the Brera Observatory in Milan, had to confess that they could not fulfil that wish, because of the lack of suitable instrumentation and buildings (Ricci 1869).

When promoting the need for a new observatory in Florence, Donati made sure that this handicap was known (Donati 1866a, 1866b). Even though he was among the pioneers of astrophysics, he planned to have an observatory that was dedicated principally to classical astronomy:

> The main purpose of Astronomy is not the investigation of the nature and essence of the heavenly bodies; these investigations are without doubts very important, but only belongs to the astronomical science indirectly ... [and] from them the astronomer is not moved much more than any other investigator of natural phenomena. But the precise purpose of Astronomy is the exact description of the movements of the stars, as those movements appear to us from Earth (Donati 1866b: 501).

While for the moment no meridian circle was available in Florence, Donati wished to prepare the conditions for the use of such an instrument: "High Observatories, like in Italy, do not conform to the times... It is necessary to have Observatories at high locations ... but in buildings as low as possible (Donati 1866b: 502)".

**3.3 A new equatorial mounting for the Amici telescope**

Donati was a strong supporter of the production of scientific instruments in Italy. There could be no real development in physics and astronomy in the country, he maintained, if there were no mechanical workshops: "Mechanical instruments are the scientists' weapons. Woe to the nation that for them must rely on foreigners!" (Donati 1868a: 353). Until the death of Amici in 1863, Donati could rely on the quality and inventiveness of his instruments. However, the small scale and domestic nature of Amici's workshop would have prevented the constructions of large devices, such as the equatorial mounting needed for the 28-cm telescope at the *Specola*. This is why Amici and Donati promoted the project of Ignazio Porro (1801-1875) to establish in Florence the "*Società Tecnomatica Italiana* ... a large plant for the construction of Precision Instruments for Astronomy,

Geodesy, the Navy, Industry, Commerce, Sciences, Arts ..." (Meschiari 2005: 23). The *Società* provided a cost estimates of 44000 liras for the construction and installation of an equatorial mounting for the Amici telescope (Ridolfi 1863a). This figure is equivalent to about 223000 € today[2].

Using this estimate, in November 1863 the Minister for Public Instruction Michele Amari (1806-1889) presented a bill to the Parliament to finance the construction of an equatorial mounting, with graduated circles, a clock drive and a dome. The bill passed and was converted into a law in February 1864 (Bianchi et al. 2012). In the meanwhile, the project of the *Società Tecnomatica Italiana* had failed to materialize and was abandoned; Instead Donati started a collaboration with the mechanic Giuseppe Poggiali (1820-1892) and established a small workshop for the production of spectroscopes and other instruments. Although an estimate of the costs was also asked to the workshop of G. and A. Merz in Münich (Merz 1864), Donati obtained authorization from the Minister to build the instrument under his own responsibility in Italy (Amari 1864). The mounting thus became the first large commitment of Poggiali's workshop, which was later named the Officina Galileo (Meschiari 2005). Donati was also authorised to visit Paris and London in order to study the latest improvements in the construction of equatorial mountings, which he did at the end of his trip to Berlin to attend the geodetic conference of 1864. In Germany, he took the opportunity to visit several observatories, which he found much better built than those in Italy (Bianchi & Galli 2014).

By the beginning of 1866, the equatorial mounting was ready, and together with the telescope tube, it was stored in a ground-level room in the Museum, awaiting installation at a suitable location (Meschiari 2005).

## 4 SEARCHING FOR A SITE

Initially, Donati and Ridolfi revived the old project for Fort Belvedere. In 1861, Ridolfi contacted the Minister for Public Instruction Francesco De Sanctis (1817-1883) and the general secretary of the Ministry, Quintino Sella (1827-1884), who liked the idea and asked for a preliminary project outline (Ridolfi 1861a). Of this project, prepared by Donati and the architect Mariano Falcini (1804-1885), we only have an indirect description by Ridolfi: a low building with three domes (most likely one for the Amici telescope and the others for auxiliary instruments) built on the site of a higher building that would have to be demolished - most likely the central building of the Fort that Antinori wanted to use. Ridolfi recommended a few solutions to reduce the costs, such as using two small domes from the *Specola*, the ruins of the old building as construction material, and avoiding Falcini's 'dream', which was to also build a Pantheon of Italian glories and a winding carriage road to access it. Concerning the Pantheon, Ridolfi (1861b) concluded that "... we already have it in the church of Santa Croce[3] … [and] the Italian Pantheon will be half-made already … " by building an observatory - thus remembering Galileo Galilei - on a Fort that reminds us the of the name of Michelangelo Buonarroti – who was responsible for the city defences of the Florence Republic in 1529. At the beginning of 1863 Ridolfi asked again to reserve the Fort for the astronomical observatory, and also for the magnetic and meteorological observatory that Donati had been asked to study when at the 1862 Turin meeting (Bianchi & Galli 2014; Ridolfi 1863b).  However, the Ministry of War replied that the Fort was needed by the military administration and was not available (Minister of Public Instruction 1863).

At the end of the year, Donati and Ridolfi reconsidered the use of the Cavaliere, which could host, if not the whole observatory, at least the dome for the Amici telescope, funding for which was being discussed by the Parliament (Donati 1863). The project was committed to the architect Fabio Nuti (Figure 9). It consisted of a circular building of 15 metres diameter with four rectangular extensions: a northern entrance, and to the east, west and south three rooms of equal size. The whole building would have a maximum extent of 30 metres in the E-W direction and occupy half of the garden. A dome with a diameter of about 7 metres would stand on top of the central room, accessed via a spiral staircase. The telescope with its basement would be installed on a pillar that rose from the ground and occupied a large part of the volume below the dome. Smaller pillars would be present under the east and west rooms to support up to four meridian instruments, and under the south room for two prime vertical instruments (but an alternative drawing with just one slit in each room exists; see Barbagli et al. 2017: 73). The space on the first floor would be divided into several small rooms, probably intended to store portable instruments that could be used on the roofs of the wings and on a balcony running all the way around the building.

Ridolfi reached a preliminary agreement about using the Cavaliere with Luigi Guglielmo de Cambray Digny (1820-1906), intendant of the royal house in Florence. However, in September 1864 it was decided to move the Capital of the Kingdom to Florence, and as part of the garden of the Royal Residence, the Palazzo Pitti, the Cavaliere was no longer available (Donati 1866a).

Both the Belvedere and the Cavaliere are within the boundaries of the city walls (that still exist nowadays in this area - see Figure 1). The city walls also delimited the extent of the Florence municipality, while the territory beyond them belonged to other administrations (in Figure 1, for example, the Torre del Gallo belonged to the municipality of Bagno a Ripoli, while the current site of the Arcetri Observatory belonged to that of Galluzzo). These limits had long been too small for the city, in particular after the move of the Capital, and they were finally enlarged at the end of July 1865 (Chiavistelli 2017). At this time, the area of Arcetri thus became part of Florence.

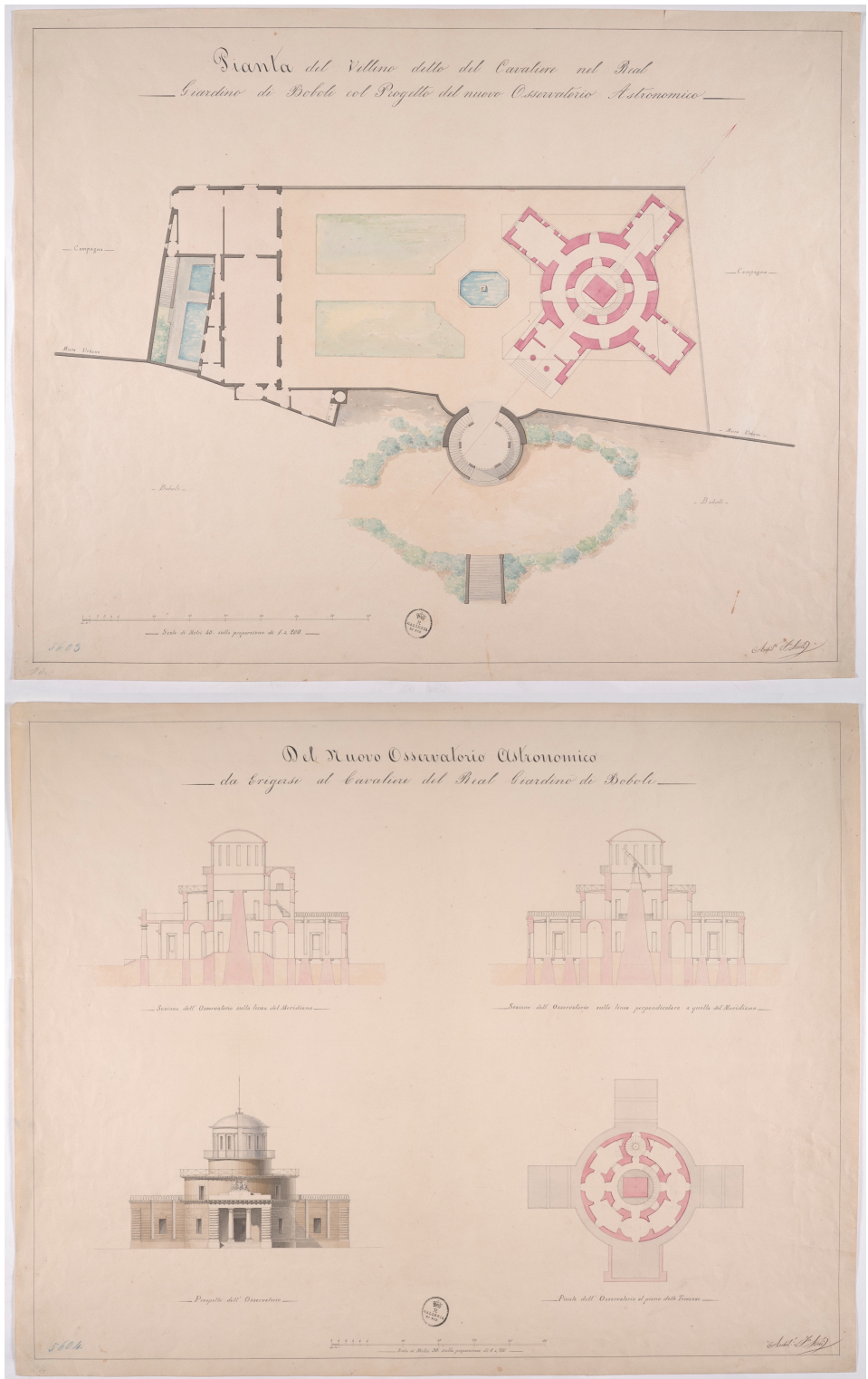

**Figure 9**: The Cavaliere Observatory project, in the Boboli Gardens. (Fabio Nuti. Gabinetto Disegni e Stampe Uffizi, Firenze; inv. 5603A, 5604A).

It is possible that the enlargement of Florence territory had a role in the final choice of the Observatory's site. After the dismissal of the Cavaliere's project, Donati and Ridolfi thought about addressing the Municipality for a proper site and a financial contribution to its building. However, Ridolfi died in the meanwhile and Donati had to wait for the appointment of the Museum's new Director, Carlo Matteucci (Figure 10). Matteucci intended moving the entire Museum to a larger building, in order to establish an advanced institute for scientific research and teaching, a normal high-school. Even before arriving in Florence, he had asked Donati to search for a "Villa of the Government or of the King which we could exchange for the Museum" (Matteucci 1865a). Eventually, a project was presented for the Villa of Poggio Imperiale (Matteucci 1866a; Figure 1). It is thus unsurprising that Donati visited, sometime before October 1865, the hill between the Villa and the Torre del Gallo. Together with the Museum's professor of Geology, Iginio Cocchi (1827-1913), and the architect Falcini, he found the location suitable because of the existence of bedrock underneath the soil (Donati 1866). Lastly, Donati was possibly aware of the ideas of Perelli, that had been published, though incompletely, in the "Civil History of Tuscany" by Zobi (1850). In fact, the words of Zobi were cited in a pamphlet supporting the project of the new observatory, written by an acquaintance of Donati (Andreucci 1868; Bianchi & Galli 2015).

**5 CLIMBING THE ARCETRI HILL**

Once the site was selected, the project for the observatory did not proceed too smoothly. With the new state incurring many expenses to complete the unification, including those for the 3$^{rd}$ War of Italian Independence in 1866, the Ministry of Public Instruction could not provide the entire costs of the new institute. Fundamental was the help of Matteucci who, until his death in 1868, was very active in seeking the cooperation of local institutions; and of de Cambray Digny (Figure 11), in his roles as Mayor of Florence, councillor of the Province, and Minister of the Finances of the Kingdom (Bianchi et al. 2013).

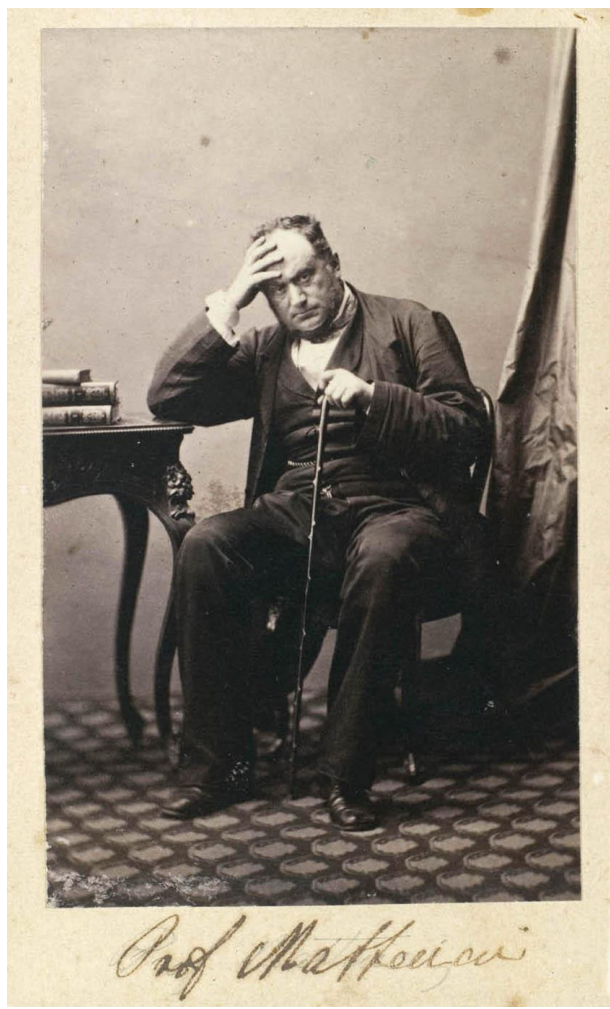
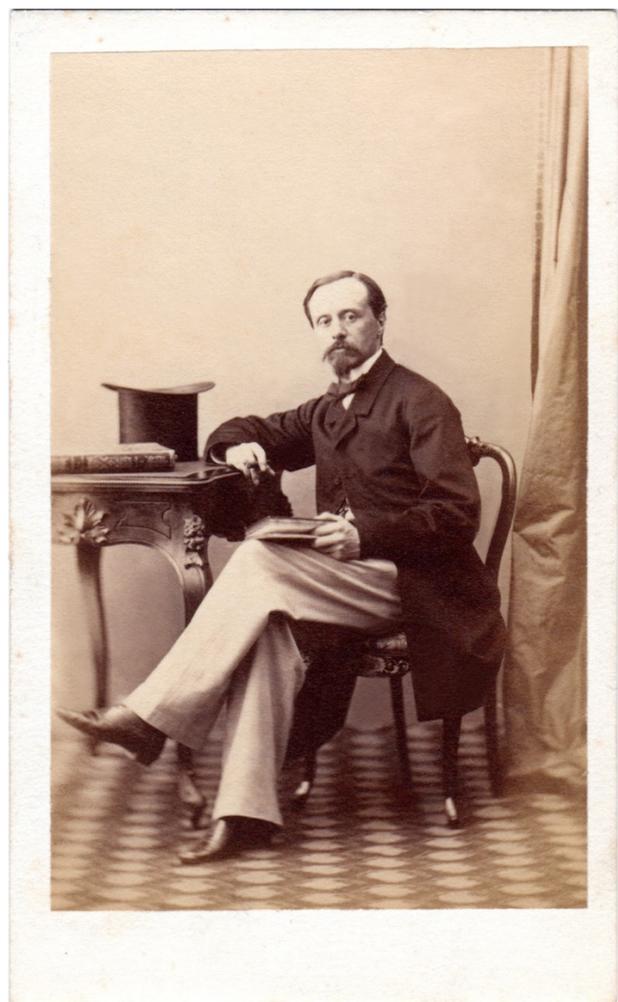

**Figure 10**: Carlo Matteucci (ca. 1860; Museo Galileo, Florence; Raccolta fotografica Cartes-de-visite raffiguranti medici e scienziati).

**Figure 11:** Luigi Guglielmo de Cambray Digny (ca. 1860; Courtesy: A. Ciabani, myarchiviostoricofotografico.com)

## 5.1 Seeking funds

The first requests were addressed to the Municipality in autumn 1865. Architectural plans and cost estimates were soon exchanged with the Mayor and the Ministry of Public Instruction, in an attempt to stimulate their support. Donati and Matteucci also tried to scare the Municipal authorities with the fear of a closure of the Observatory if the current situation persisted. While at the same time renouncing to the ambitious plan to move the Museum to the Villa of Poggio Imperiale (occupied by a school), on the eve of the vote by the Municipal Council Matteucci urged de Cambray Digny:

> The important thing is to have an observatory and believe you that if a new observatory is not constructed, the government will feel the need to abandon that of the Museum because it is absolutely useless and such that it cannot even receive the instruments we have. Thus, Florence will remain without an observatory... The council ... must make just one decision, that is, if Florence should have the first astronomical observatory of Italy, yes or no: if the council gives a subsidy to the government, Florence will have this leading observatory ... (Matteucci 1866b).

On 9 June 1866 the Municipal Council voted for a subsidy of 30000 liras, providing that both the Province and the Government also gave contribution (*Atti del Consiglio Comunale di Firenze*, 1872:559-562).

In the meanwhile, Matteucci asked Donati to promote the project to a larger audience. In November 1866, the astronomer gave the inaugural lecture of the Institute, stressing the inadequacies of the existing Italian observatories and presenting his proposal "of erecting in Florence a new Observatory that could serve the modern needs of astronomy" (Gazzetta Ufficiale del Regno d'Italia 1866; passage from Donati 1868b). Donati (1866b) had presented his arguments earlier in March in the first issue of the *Nuova Antologia*, a new monthly review of "Letters, Sciences and Arts", and in autumn 1866 he also published the memoirs he had presented to the Municipality in 1865 (Donati 1866a). Donati (*ibid.*) also included a memoir sent to the Council of the Province in September 1866. The Council approved the co-financing of the new observatory at the Assembly meeting of 27 November 1866, where de Cambray Digny intervened to make sure that the Province would contribute with the same amount as the Municipality (*Atti del Consiglio Provinciale di Firenze* 1867:235-239).

A final contribution came from the Royal House. When Victor Emmanuel II transferred to Palazzo Pitti, he chose as private apartments a part of the palace named Palazzina della Meridiana (Rensi 2015), very close to the *Specola* (Figure 1). From the observing room, it was possible to look into the apartments and, apparently, into the King's bedroom (Maison 2003). In order to remove this 'annoyance', funds were promised in exchange for the upper part of the tower (Matteucci 1865b). While the observatory was still to be transferred, the observing room of the *Specola* thus became part to the Civil List (Gazzetta Ufficiale 1868). Later the Royal House offered a contribution of 15000 liras for the new observatory (Relazione 1869).

## 5.2 The project

Plans for the Observatory were presented by the architect Falcini (1865) in November 1865. The Observatory was to sit on a hilltop belonging to two private land-owners and have a northern access via a short winding road that connected to the street. Soon after, however, it was realised that some of the expropriation expenses could be saved by using nearby land owned by the State (the Podere della Cappella, Chapel's farm; Bianchi 2017). Thus, in August 1866, modified plans were presented with a longer access road from the South, which passed through the Podere della Cappella (Minister of Public Instruction 1867).

The cost of the project was estimated at about 115000 liras, more than half of which was covered by the contributions from the Municipality and from the Province (that of the Royal House was yet to come). Matteucci (1867a) then appealed to national pride in order to encourage his superiors:

> ... despite the financial straits of our budget, it will not be difficult for the King's government to obtain from the Parliament a sum which is very small and that is destined to preserve illustrious traditions and to keep astronomical studies in Italy not so far from the highest destinations to which they have already ascended and continue to rise among the other civilized nations of Europe. And the most recent example of the national observatory built in Switzerland with an even greater expense than that foreseen by us will have to comfort the King's government and the Italian Parliament not to always remain below much smaller states in promoting scientific studies and good and rigorous observational methods.

In this letter, Matteucci was most likely referring to the Swiss Federal Observatory that was built in Zurich between 1862 and 1864 (Friedli & Keller 1993).

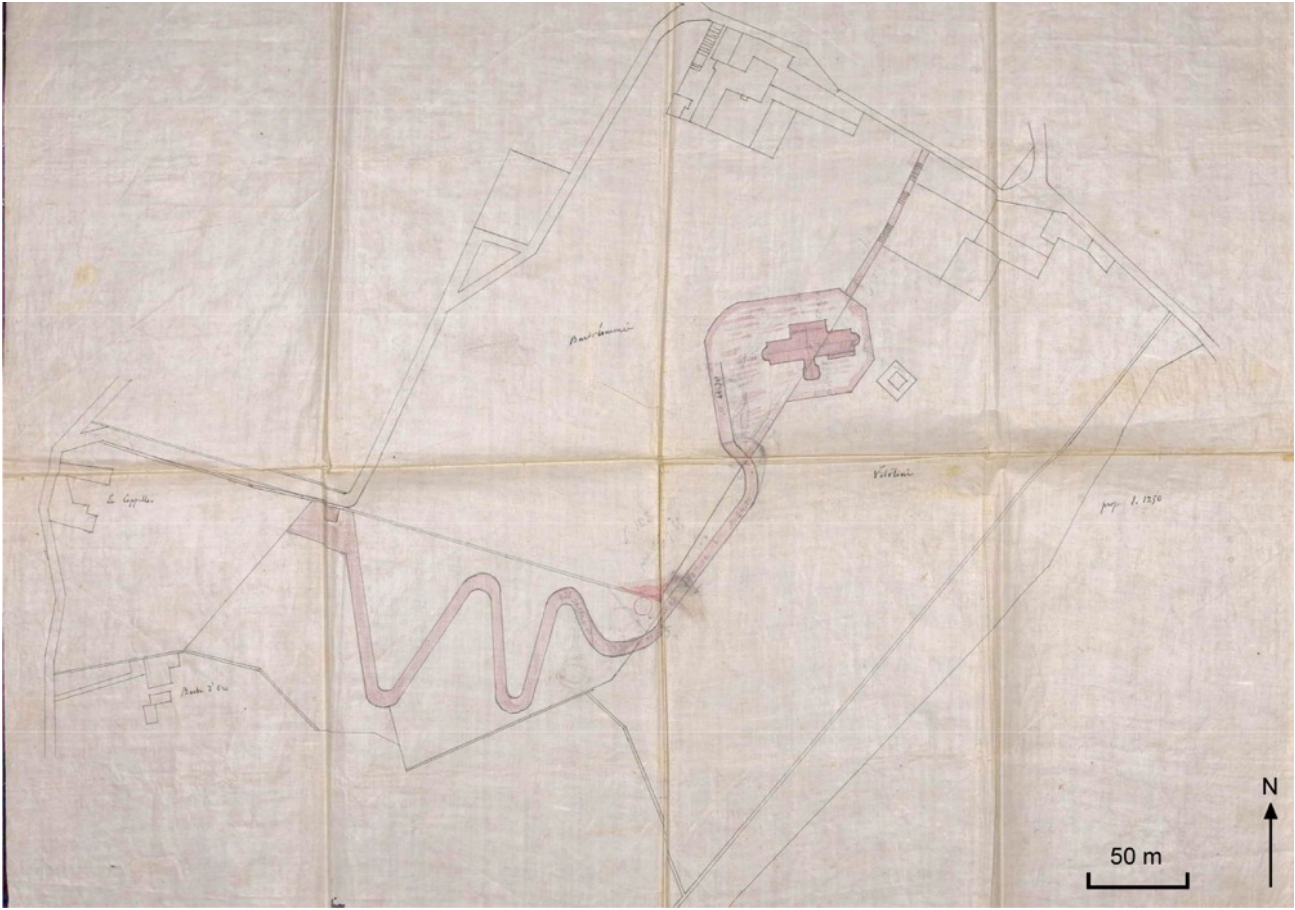

**Figure 12:** A cadastral map with the first project of the observatory (Archivio Storico Università, Florence; Fondo Carteggio della Soprintendenza, 1894/473). The map is undated but was likely made early in summer 1867. The red circle, almost in the centre of the map, represents the final position of the provisional pavilion for the Amici telescope (Sect. 6.1); two, faint, pencil ones probably show the first choices.

None of the drawings of this first project has been found and the original layout of the building has to be guessed from the written descriptions and from a sketch on a cadastral map (Figure 12). The building had an entrance gallery from the north; an east wing with the apartments for the astronomer, with a few extra rooms protruding to the south; and a west wing for the meridian instruments. Two small domes were positioned on the east and west ends, while the larger dome for the Amici telescope was on a building extension to the south. The Superior Council for Public Works recommended a few changes in August 1867. Among them, they required a major building symmetry: only the northern entrance should be a forepart to the east and west wings; the larger dome should be positioned over the centre of the building and a few extra rooms protruding from the east wing should be removed (Ferrucci 1867).

The final, corrected, project was ready by the end of 1867 (Matteucci 1867b). To this project belongs a single surviving drawing showing the façade of the building and the E-W section of its central part (Figure 13). The section shows the mechanisms for a revolving, cylindrical, dome and the telescope basement, above a pillar rising from the building's foundations. Curiously, it does not yet show the slits for the meridian instruments.

**5.3 The law for the observatory**

In 1868, controversial affairs in Italian politics were the concession of the tobacco monopoly to a private society and the introduction of a tax on milling, two measures presented by de Cambray Digny as Minister of Finances and intended to consolidate the State budget. At the autumn start of Parliament, Donati urged de Cambray Digny to present a long-awaited bill about the Observatory:

> Excuse me for pity's sake if I dare to disturb ... the most natural course of your occupation, coming to talk to you not about Tobaccos, or Milling, but Astronomy! (Donati 1868c)

> ... Astronomy persecuted you first at the Town Hall, then at the Province, and now at the Ministry! ... You, for having already done so much, will certainly not refuse to do now, what is missing for the completion of the work! (Donati 1868d)

The bill was presented to the House of Deputies by de Cambray Digny on 21 January 1869. The Parliamentary Committee for the bill's evaluation promoted it using the same arguments developed previously with the Municipality and the Province (Berti 1869). After finally managing to obtain the contribution of the Royal House, the Committee recommended building the Observatory because of its low cost: out of a final estimate of about 106820 liras (slightly reduced after the changes to the project, and equivalent today to about 520000 €), 75000 had been promised by the Municipality, Province and Royal House and about 27500 had been allocated already in the budget by the Ministry of Public Instruction in 1868. Members of the House were thus required to allow a further expense of just a little over 4000 liras. The bill was also modified to include the whole of Podere della Cappella into the Observatory's terrain. The House of Deputies approved the bill on 25 May 1869 and passed it to the Senate. However, in June the Senate was closed by the King and all bills under discussion were cancelled (Bianchi et al. 2013).

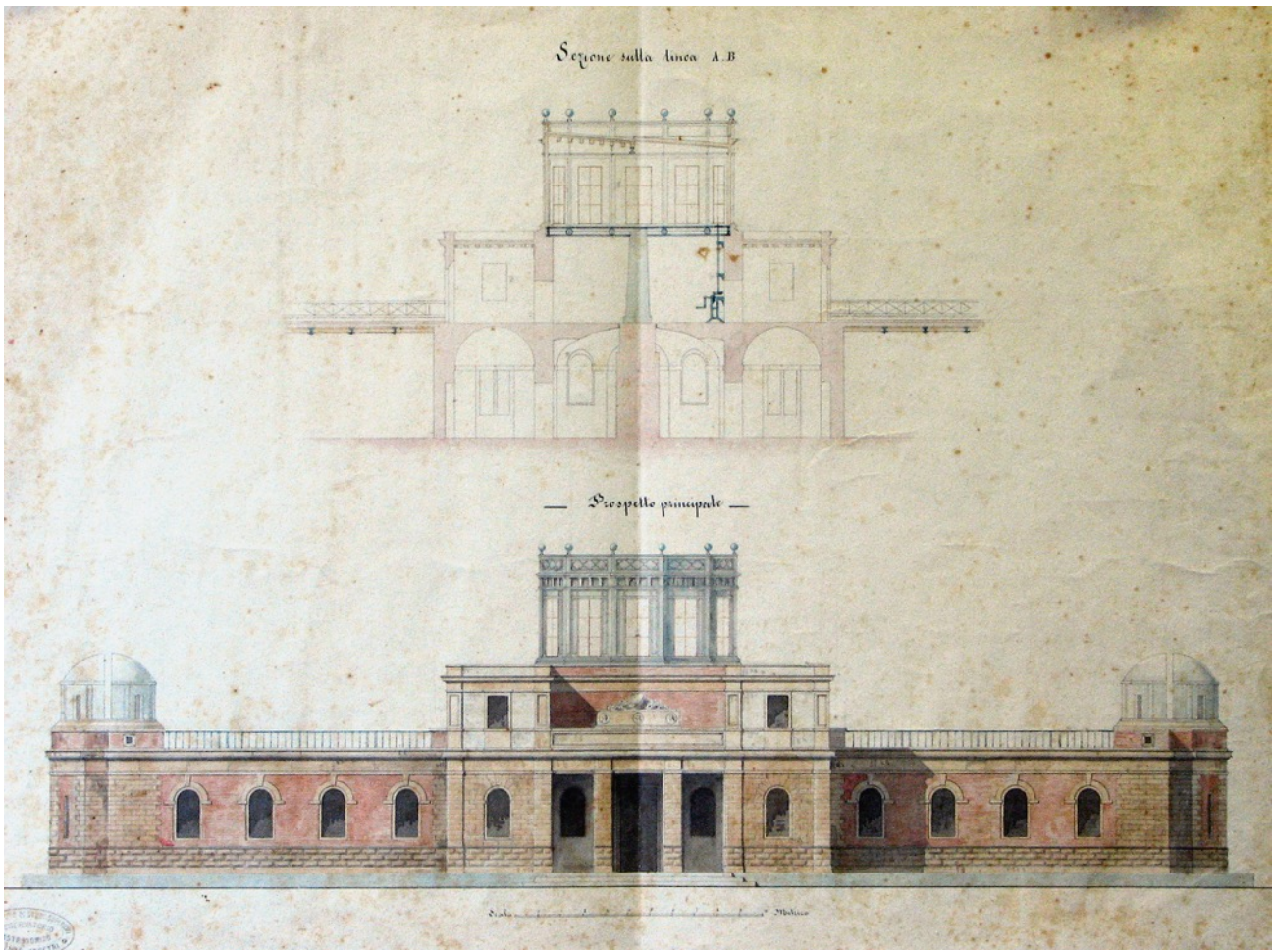

**Figure 13:** The only surviving drawing of the project of the Observatory, ca. 1867 (Archivio Storico INAF-Osservatorio Astrofisico di Arcetri).

To avoid delays, the bill was converted into a Royal Decree, issued on 23 September (Relazione 1869). However, the Parliamentary approval of the Decree did not proceed, because of political and historical events, among which where the conquest of Rome and the final transfer there of the Capital and of the Parliament. The Ministry of Public Instruction later decided to put a further sum of 30000 liras in its budget for 1871 (Cipolletti 1872; Donati 1873a).

**6 THE BUILDING OF THE OBSERVATORY**

The building on Arcetri hill started just when the bill was being presented to the Parliament, early in 1869. The construction proceeded in two steps, balancing at a time the desire of Donati to speed up the process, and the delays due to bureaucracy, to projects' modifications, and to parliamentary works. First, the Amici telescope was provisionally installed in Arcetri, using in part the leftovers of the funding for its mount, and then the main building and the final installation of the telescope were completed. Because of this, the Observatory has the peculiarity of having been inaugurated twice (Bianchi et al. 2013).

**6.1 The provisional installation of the Amici telescope**

After the mounting for the Amici telescope was ready, Donati searched for companies that could build the dome. Cost estimates for an iron cylindrical dome with a spherical cap were asked in June 1866 (Donati 1866c). Later it was decided to build a wooden dome with iron frame and movements, and a copper roof (Matteucci 1867c). The dome, about 9.5-m across, had a polygonal cross-section with twelve sides, one occupied by a slit and six by glass windows. All parts of the dome were made in Florence and completed at the beginning of 1868 (Falcini 1868).

At the same time, permission was asked to build a pavilion on the Podere della Cappella (Matteucci 1867d). The pavilion was intended to host instruments for the measurement of the terrestrial magnetic field. However, its circular walls would be used first as a base for the provisional installation of the dome and the telescope, while awaiting the construction of the Observatory – on land that still had to be expropriated (Bianchi et al. 2012; 2013). The works for the pavilion were authorized at the end of summer 1868, and a road was built to access the higher part of the Podere della Cappella, the closest place to the top of the hill. The dome with the telescope installed inside was ready by June 1869 (Donati 1869a).

It is likely that Donati sped up the building in order to have it ready by the end of September 1869, when the Permanent Commission of the *Europäische Gradmessung*, together with several Italian astronomers, met in Florence. On Sunday 26 September, geodesists and astronomers converged on Arcetri along with other Professors of the Institute and authorities (Figure 14). The Italian Government was represented at the highest level, by the Prime Minister Luigi Federico Menabrea (1809-1896), the Minister of Public Instruction Angelo Bargoni (1829-1901) and the Minister for the Royal House Filippo Gualterio (1819-1874). The meeting was solemn, as it should have been for the inauguration of a National Observatory in the Capital of the Kingdom. According to Donati (1869b), that was the day that should rightly be "… considered as the first from which it must be intended that the history of the New Florentine Observatory begins" (Donati 1869b).

**6.2 The main building**

The construction of the main building had to wait for the expropriation contracts, which were signed in the spring of 1870. The road was then completed from the provisional location of the dome up to the hilltop. Just before the start of the work on the new building another curious hindrance emerged: some inhabitants of nearby villas expressed concerns about the loss of privacy of ladies in their rooms. Donati had to explain that "however similar to celestial bodies the ladies might be, such telescopes as used at the observatory could not be used towards objects so removed from the heavens *in distance!*" (Baldelli 1870).

In September 1870 construction was assigned to the contractor, Carlo Berti, who had already made the road and the masonry of the pavilion. A few problems emerged between Donati and the architect Falcini. For scientific reason, the astronomer required a few changes to the project: the building was moved to the south side of the hilltop; the meridian direction was recalculated and corrected, which required a modification to the foundation walls after they had already been laid (Boccini et al. 1881; Falcini 1873; 1879). In February 1871, Donati (1871a) reported: "All the foundations are already completed and now the building starts to emerge from the ground, and I hope it will be finished soon". By this time, other project modifications had probably been decided: the move of the entrance to the south façade; a long, additional, external stairway on that side; an increase in the height of the building; and the removal of the central pillar for the telescope, whose weight now had to be sustained by the dome vault of the central room on the ground floor. All of these changes caused a rise in the final costs of the building (Falcini 1873).

By the end of summer of 1871, the Municipality and Province gave, at least in part, the financial contributions they had promised. Donati also asked the local authorities to express their satisfaction with the works, in order to justify the new unforeseen additional expenses. After a visit of two Councillors (one of them being the ubiquitous de Cambray Digny), the City Council stated that " … this Institute will be the only one in Italy reaching the degree of perfection that is now deemed necessary …." and recommended that the Minister of Public Instruction, Cesare Correnti (1815-1888) increase the budget to ensure the completion of the work (*Resolution* 1871).

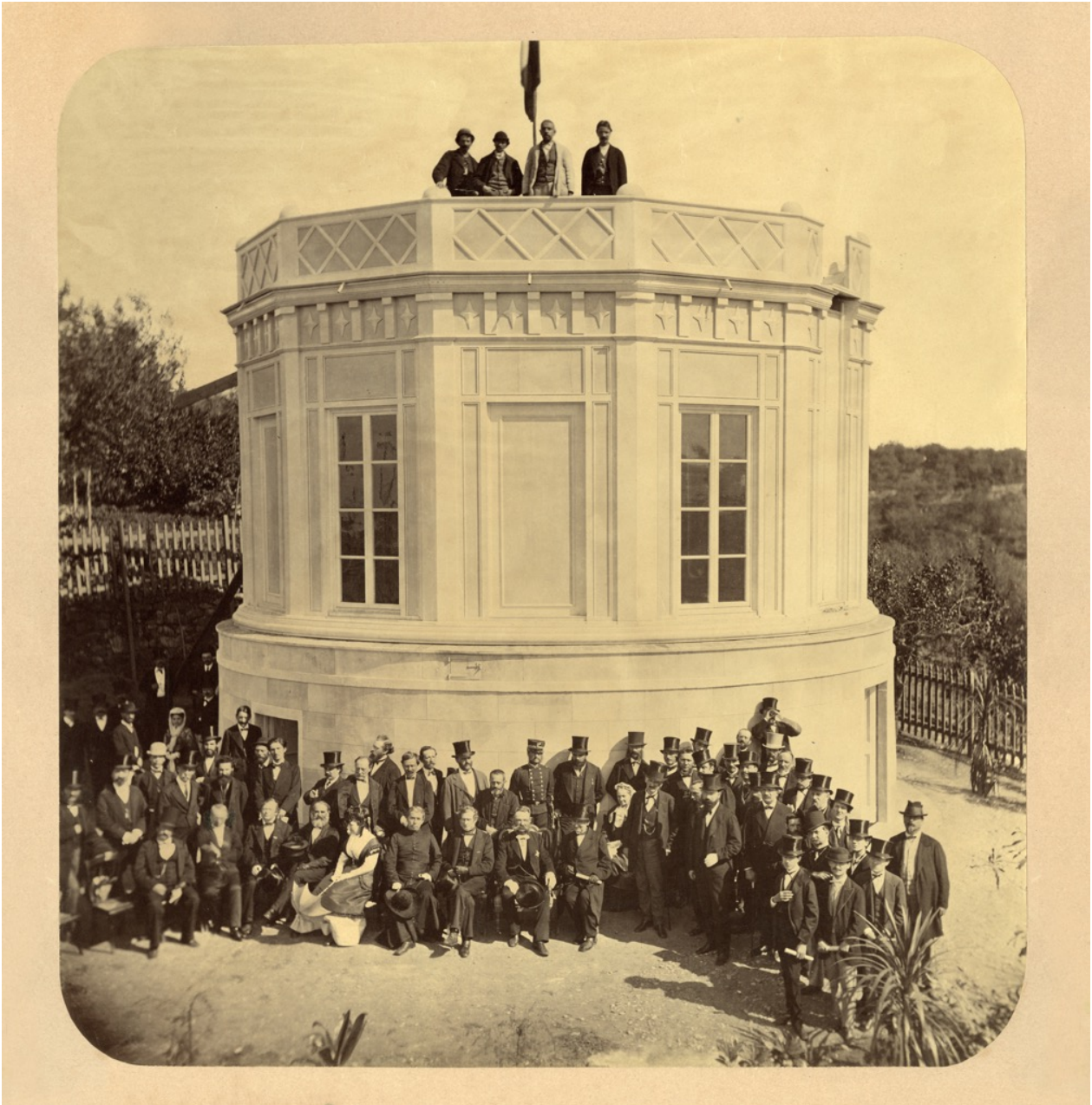

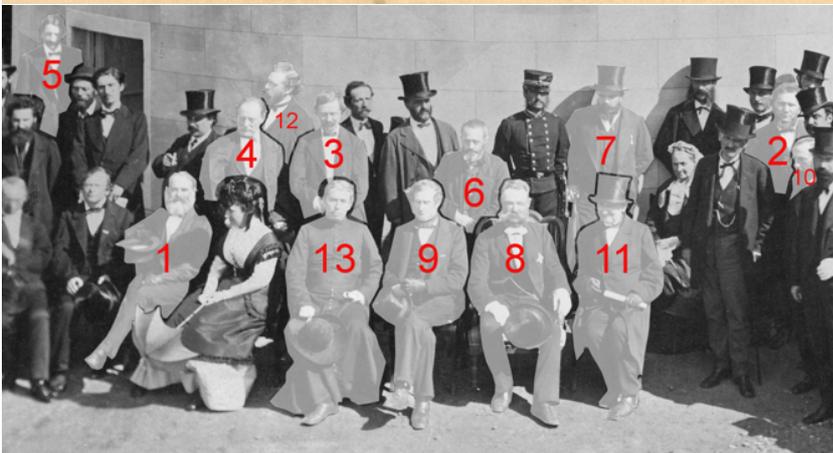

Identified:
1. Baeyer, Johann Jacob (1794-1885) [likely]
2. von Bauernfeind, Carl Maximilian (1818-1894)
3. Bruhns, Carl Kristian (1830-1881)
4. Cacciatore, Gaetano (1814-1889)
5. Cipolletti, Domenico (1840-1874)
6. de Gasparis, Annibale (1819-1892)
7. Donati, Giovan Battista (1826-1873)
8. von Fligely, August (1810-1879)
9. Kayser, Frederick (1808-1872)
10. Lorenzoni, Giuseppe (1843-1914)
11. Santini, Giovanni (1787-1877)
12. Schering, Ernst Christian Julius (1833-1897)
13. Secchi, Angelo S. J. (1818-1878)

Other participants:
Forsch, Eduard
Hirsch, Adolph (1830-1901)
Ibáñez e Ibáñez de Ibero, Carlos (1825-1891)
Peters, Christian August Friederich (1806-1880)
Ricci, Giovanni (1813-1892)
Schiaparelli, Giovanni Virginio (1835-1910)
Schiavoni, Federigo (1810-1894)

**Figure 14:** The first inauguration of the Arcetri Observatory, on 26 september 1869 (F.lli Alinari, Firenze. Archivio Storico INAF-Osservatorio Astronomico di Padova). The bottom key includes the names of the astronomers and geodesists attending the ceremony, some of whom have been identified in the picture. For more details, see Bianchi et al. (2013).

By the autumn of that year Donati felt discouraged. He had already handed over the main room of the old observatory to the Royal House, while the new observatory building was still the domain of the masons, but just a few of them, since "the contractor had lost confidence knowing that he cannot finish the building with the allocated funds and he fears with reason that he will have to wait indefinitely before being paid" (Donati 1871b). Also, the Ministry was deaf to his requests, having left Florence for the new Capital, Rome, before settling the problems of the funding. Indeed, the move of the Capital took some momentum away from the Arcetri Observatory project. In a discussion in the Senate on 20 December 1871, General Nino Bixio (1821-1873), a friend of Giuseppe Garibaldi and one of the organizers of the 1860 Expedition of the Thousand, asked the Minister of Public Instruction if something was planned to transform the Roman Observatory into an observatory worthy of a Capital. The Observatory Bixio was referring to was that of the Collegio Romano run by Secchi and dedicated mainly to the 'novelty' of astrophysics, while he judged of more utility works relating to ephemeris calculation and geographical determinations, such as "those of the Observatories of Greenwich, Paris, Washington, Madrid and others." Minister Correnti replied:

> The standard Observatory at this moment is the Florentine Observatory, for which, as the Senate knows, a considerable expense was made with the help of the civil list, of the Province and of the city of Florence. The Observatory of Florence located on the hill of Arcetri, in an excellent position, is the only one in Italy that can compete with the principal observatories of the great nations (*Rendiconti* 1872:56).

However, Correnti concluded that in the near future it might be useful to have an observatory in the Capital. Apparently, the Minister was unaware that a public facility already existed in Rome in the form of Campidoglio, which belonged to the University.

Despite all odds, the work at Arcetri proceeded. The Amici telescope and its dome were removed from the pavilion and installed in the main building, which by the summer of 1872 was almost complete (Figure 15). Donati scheduled a new ceremony for October: if the first event of 1869 was to be considered as the laying of the "first stone", this second inauguration was to celebrate the "placement of the last stone of the Observatory of Arcetri" (Donati 1872a). The astronomer wished to have the same attendance as for the first inauguration, so invitations were sent to various Italian astronomers and physicists, while the Museum director, the botanist Filippo Parlatore (1816-1877), also invited the attendee of the conference of the International Metre Commission in Paris (the invitation was read at the October 4 session; Commission Internationale du metre 1872: 60). The ceremony took place on Sunday 27 October (Figure 16). Though the attendance was large, only a few international guests were present, probably because of problems the trains due to poor weather (the event itself had been postponed for a week). But among them who did attend was France's Camille Flammarion (1842-1925). Donati himself was missing, because he broke his leg the day before. Other notable absentees were the Minister of Public Instruction Antonio Scialoja (1817-1877) and any other major Government representative: by this time, Arcetri had definitely lost its appeal of a National Observatory (Bianchi et al. 2013).

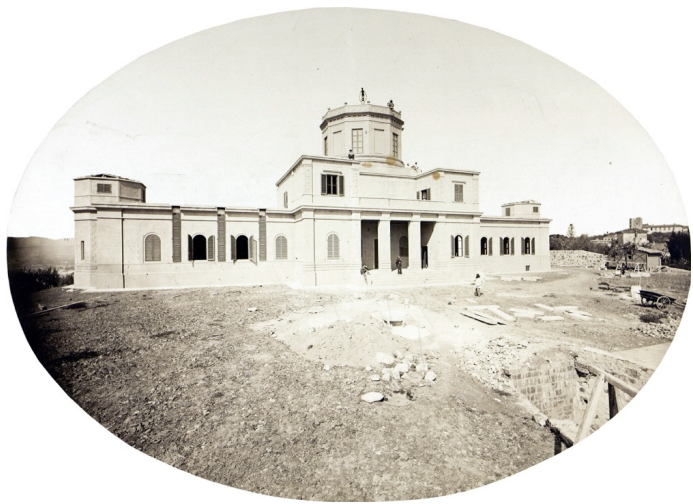

**Figure 15**. Finishing the construction of Arcetri Observatory, presumably in summer 1872 (F.lli Alinari, Firenze. Museo Galileo, Firenze).

**6.3 Early science from Arcetri**

Donati must have tried to use the Amici telescope as soon as it was installed in Arcetri in its provisional position. He also had a wooden cabin built next to the telescope, in order to facilitate its nightly use and be able to follow the construction works (Bianchi et al. 2013). However, the first documented observations made from Arcetri Observatory were not done with the refractor, but with a Repsold transit instrument belonging to the Italian Geodetic Committee. Stellar transits were observed in October and November 1869, both at Arcetri and at the Adriatic port of Ancona, in order to measure the difference in longitude between the two locations. The observations, which involved the use of a telegraphic line brought to Arcetri for that purpose, were part of the *Europäische Gradmessung* operations (Donati 1871c).

The first (and only) documented use of the Amici telescope in its new pavilion is dated June to August 1871. Donati (and likely his astronomy Assistant Domenico Cipolletti, 1840-1874) observed Comet C/1871 L1, discovered by Wilhelm Tempel (1821-1889) from the Brera Observatory in Milan (Donati 1871d). That same August, the astronomers also counted meteors of the Perseid shower (Donati 1871e).

The Amici telescope must certainly have been used for other observations when Donati stayed at Arcetri overnight. On the evening of 24 October 1870, he observed a magnificent aurora "from the heights of the new observatory of Arcetri, and not disturbed therefore by the light in the streets of the city, which is always injurious to any celestial observation." (Donati 1870). Another aurora was observed on 18 April 1871, when Donati was preparing to search for Comet C/1871 G1, most likely from Arcetri (Donati 1871f). A third bright aurora occurred on 4 February 1872, which Donati saw from the city (Donati 1872b). In these phenomena and their connection to terrestrial magnetism and solar activity, Donati saw the effects of a 'cosmic meteorology', one of the first definitions of Space Weather (Cade & Chan-Park 2015).

While the Amici telescope was being moved from the provisional pavilion to the main Observatory building, Donati installed the Fraunhofer telescope at Arcetri (Donati 1871g). The old refractor had been equipped with a new equatorial mounting and a spectroscope, made by the Officina Galileo, and was used in Sicily during the solar eclipse of 22 December 1870. At Arcetri, the telescope was placed in a small dome, which derived from the old *Specola*, on the south-west terrace in front of the Observatory. The only instrument available for use in February 1872, it was used to monitor solar activity and observe the photospheric spectrum, on the days before and after the auroral display (Donati 1872b). A new high-dispersion 25-prisms spectroscope attached to the Fraunhofer telescope allowed to observe in April the inversion of the H$\alpha$ line on sunspots (Donati 1872c); in June, the chromosphere in the D$_3$ Fraunhofer line, later known to be due to Helium (Bianchi et al. 2016; Donati 1872d).

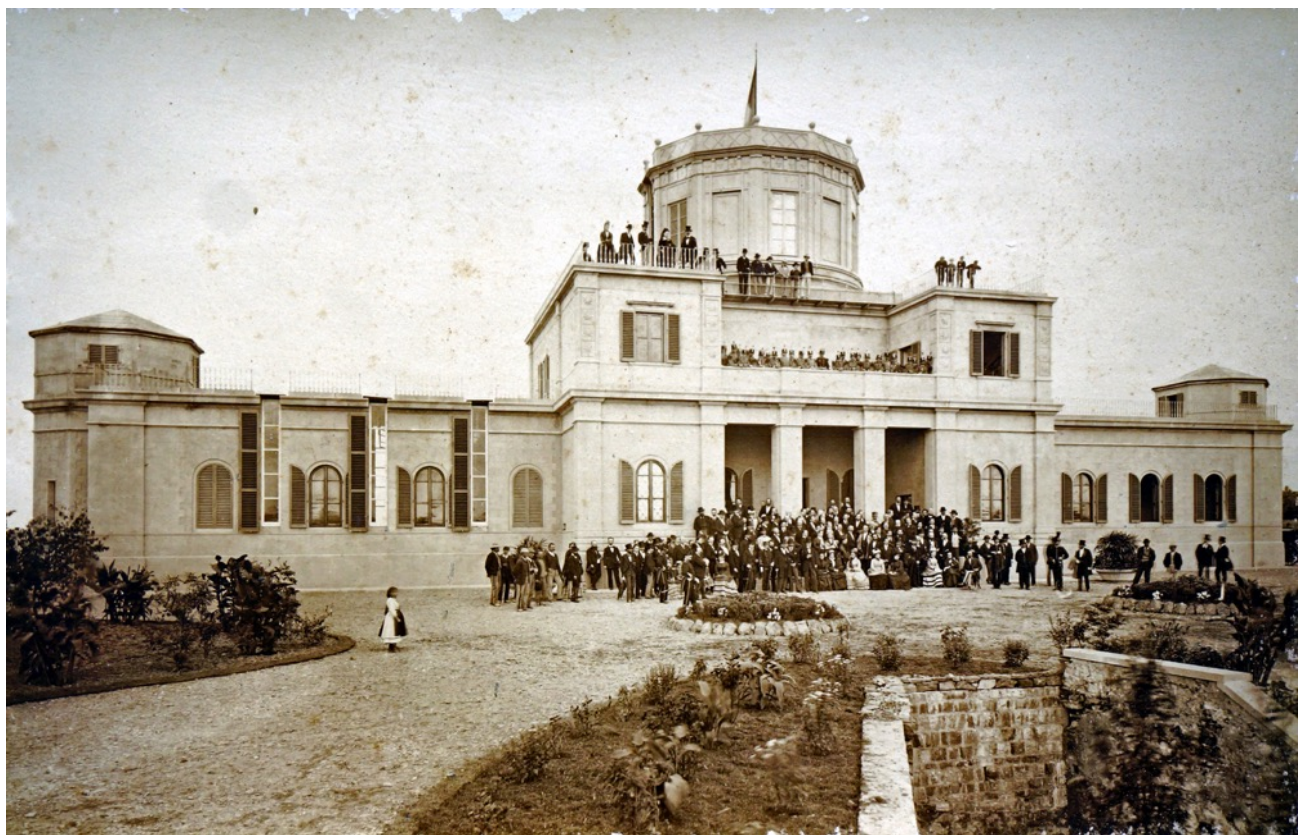

**Figure 16**. The second inauguration of Arcetri Observatory, on 27 October 1872 (F.lli Alinari, Firenze. Archivio Storico, INAF-Osservatorio Astronomico di Monte Porzio, Roma).

## 7 STRUVE, PULKOVO AND ARCETRI

In October 1871 Arcetri Observatory was visited by Otto Wilhelm Struve (1819-1905; Figure 17). The German-Russian astronomer was one of the most authoritative figures in 19[th] century astronomy. In 1862 he became Director of the Pulkovo Observatory near St. Petersburg, one of the best built and equipped astronomical observatories of the time, founded by his father Wilhelm in 1839 (Batten 1988). Here is how Donati remembered Struve's visit in the speech for the 1872 inauguration:

> Here came, among others, Struve, the illustrious Russian astronomer who, in the several reports his Government asked him on the state of practical astronomy in the numerous countries he visited, never ceased to deplore the miserable conditions of Italian observatories, and whose authoritative voice, having been heard by our own Government, certainly had not a minor influence on the conception and realization of this Observatory. Struve wisely directs the large observatory of Pulkova, which is a truly and splendid scientific royal palace of which he is the prince (Donati 1872e: 5).

Donati believed that Struve provided substantial impetus a big to the Arcetri project. The Italian astronomer hoped that his Russian colleague could be present at the 1872 ceremony, when Struve would have been a guest of honour and bestowed with a high decoration. However, Struve could not be present (Bianchi & Galli 2015).

Among Struve's "several reports", one describes a tour of almost all Italian Observatories - including the *Specola* of Florence - that had taken place a decade before, in Autumn 1863. Struve found all the Italian observatories poorly manned, equipped and badly built; he found that no one of them followed "those rules ... which already for more than half a century have been recognized by the scientific world as essential for the success and exactness of observations" (Bianchi & Galli 2015: 217). At the end of the visit, Struve met the Minister of Public Instruction of the Kingdom of Italy and presented his recommendations for Italian Observatories:

> ... instead of the large number of these establishments spread across the Kingdom and almost none of which were provided with sufficient means to the current requirements of science, it would be useful to have only a small number that are well-organized and suitably-equipped ... (Bianchi & Galli 2015: 226).

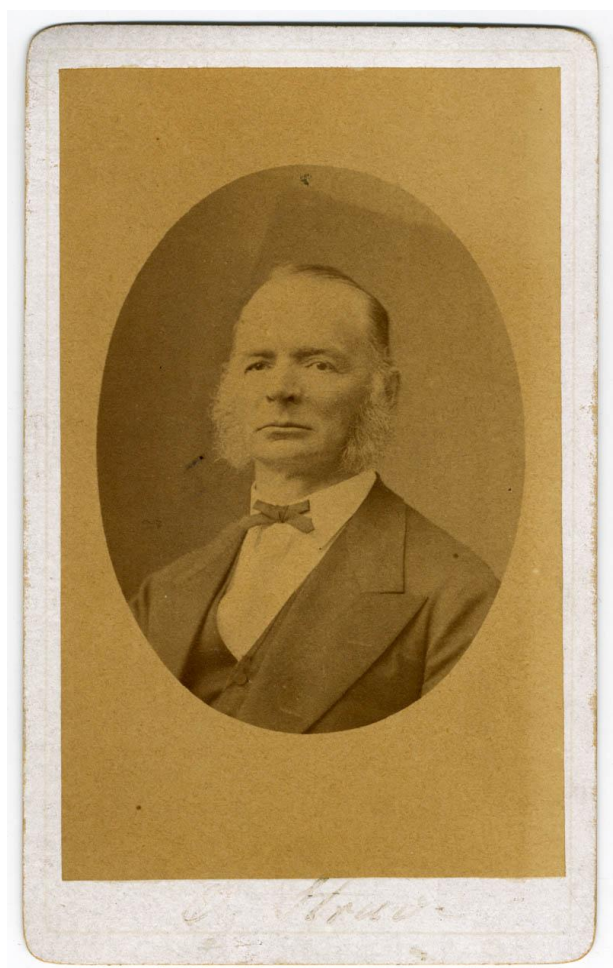

**Figure 17:** Otto Wilhelm Struve (ca. 1865; Museo Galileo, Florence; Raccolta fotografica Cartes-de-visite raffiguranti medici e scienziati).

Back home, Struve reported to his own Minister and his account was published in Russian in 1864, apparently unnoticed to Italian astronomers.

In Autumn 1867 Donati and Struve met in Berlin at the conference of the *Europäische Gradmessung*. It was probably on this occasion that Donati learned of the report. Struve sent him a handwritten copy of the report, translated into French. Donati immediately started to use Struve's report to support the building of the new Observatory in Florence (Donati 1868b). References to the report are also present in the Parliamentary discussions on the bill for the Observatory and on the budget of the Ministry for Public Instruction (Bianchi & Galli 2014; 2015). After his 1871 visit to Arcetri, Struve sent Donati a letter expressing his satisfaction with the new Observatory, exhorting the authorities to complete it and provide the necessary means. This document also was used by Donati, who passed it on to the Ministry (Bianchi & Galli 2015).

Under the Directorships of the Struves, father and son, Pulkovo became a model for the building of several observatories worldwide (Wolfschmidt 2009). It also became a source of inspiration for the Arcetri project. The east-west layout of the Florentine building is reminiscent of the Russian Observatory, although the design was not unique to Pulkovo (Castro Tirado & Castro-Tirado 2019) – for example, in Italy it characterises the Capodimonte Observatory in Naples that was built from 1812 (Pigatto 2012). Donati and Falcini must have used the *Description de l'observatoire astronomique central de Poulkova* (Struve 1845) as a sort of handbook, to search for ideas and technical solutions, just as Amici and Antinori had done earlier with Dorpat Observatory (Struve 1825). Indeed, Donati (1864b) asked Struve about details of the clock drive of the Merz refractor, as outline on page 185 in the *Description*. In Falcini's project, the roof of the meridian hall was supported by pillars and arches on each side of the three vertical slits, as "in the drawings of the Pulkovo Observatory in Russia which is reputed to be one of the best built" (Comparini 1878). There were plans at Arcetri to install a transit instrument on the prime vertical, as Wilhelm Struve did in Pulkovo (Cipolletti 1872; Donati 1869b). There is also the suggestion of a further link: in 1866 a wood-seller from Paris, Alphonse Thibaut, was paid for preliminary drawings of the dome for the Amici telescope (*Conto del R. Museo* 1866). Was he related to the "M. Thibaut, former stage technician", who built the domes for Pulkovo (Struve 1845: 36)?

**8 COMPLETING THE OBSERVATORY**

Despite the 1872 inauguration, the Observatory was still incomplete, both with the building and with the instrumentation. Donati was still recovering from his broken leg and felt discouraged. He confessed to Schiaparelli: "If I had an enemy, I would wish him to be an astronomer, and that it would come into his head, as it came to mine, the idea of making a new observatory" (Donati 1872f). Nevertheless, Donati persevered. At the end of the year, he proposed that the Municipality build a pedestrian access from the north, connecting the Observatory to the new *Viale dei Colli* (Figure 1), an avenue made by the architect Giuseppe Poggi (1811-1901) and part of a plan of modernization started when Florence was the Italian Capital (Poggi 1882). Perhaps, by connecting it to the *Viale*, which was already a popular promenade, Donati hoped to gain visibility (and support) for the Observatory. However, the project was not implemented, and even nowadays the northern access is limited to a narrow passageway that leads to the nearest street.

In the end the building had cost 197000 liras, almost twice as much as the original estimate (Donati 1873a;b). The rise was also due to inflation between 1869 and 1873, since the cost is only 1.5 times more than the original, when converted to current values (about 770000 €). Most of the extra expense had been sustained by the contractor Berti, who demanded to be paid. Donati thought about solving the issue by proposing an increase in the funding from the Ministry, to be supported by a refurbishment of the law that had never been approved (Donati 1873a). The cost of the missing instruments, instead, was to be covered by the Municipality and the Province, that in June 1872 had signed an agreement with the Minister of Public Education to fund the needs of the Institute of Superior Studies, to which the Observatory belonged. For the Observatory, 87000 liras of extraordinary expenses had been allocated (*Riordinamento* 1872). Donati tried to convince the authorities that the total cost, including the building and the instruments, actually was small, when compared to those of foreign observatories. For example, Pulkovo - the unreachable reference at the top of the scale - had cost "about 8 and a half million of current liras", although Donati recognized that it was "the richest of all that exists, and truly its example dismays". Just to mention the more recent ones, the new Leiden Observatory cost 3 millions of francs; the Dudley Observatory in Albany, NY, half a million liras; and the Observatory of Lepzig, 200000 liras - a cost comparable to what was needed by Arcetri, although the German observatory was "more modest than the Florentine Observatory" and was financed by "the small Kingdom of Saxony" (Donati 1873c). The Kingdom of Italy, Donati implied, could certainly afford to spend more for the sake of Science.

When asking for new instruments, Donati maintained his original ideas mentioned above, that the Observatory had to be dedicated to classical astronomy:

> The New Observatory, for its position and construction, must be especially directed towards the astronomical observations that are called <u>fundamental</u>, that is the determination of the positions of celestial bodies.
>
> This purpose, although less universally understood, and, I will say, less seductive than those other studies that target the physical constitution of the stars, is still the most important of Astronomy (Donati 1873b).

These are indeed strange words from the mouth of a pioneer of astrophysics! Nevertheless, the largest expense that Donati asked was that for a great meridian circle "that the Observatory of Florence has never had", with a circle diameter of at least 1 metre and possibly made by the Repsold firm in Hamburg. Clearly, Donati considered that this instrument was too complex to be constructed by the Officina Galileo of Florence. For the rest, he suggested modifications to the old Sisson transit instrument and required just a spectroscope for the Amici refractor on its equatorial mounting, which was considered complete (omitting that it still did not have graduated circles and a clock drive). He also asked for the small domes on the east and west ends of the Observatory building (in Figure 15 and 16 they are covered by provisional roofs); a few small buildings for laboratories, workshops, the gatekeeper and assistants; and the completion of the pavilion to install the self-recording magnetic instruments he had bought earlier in London - which would be used for research on 'cosmic meteorology' (Bianchi et al. 2011; Cipolletti 1872; Figure 18).

Other problems were already emerging. Donati had asked the builder to follow his own design for the support of the meridian slits, not Falcini's original projects. Already in September 1872, the roof proved unable to shelter the meridian hall from heavy rains and a modification was needed (Falcini 1879). Donati was certainly ready to face this and other difficulties, but unfortunately he died suddenly on 20 September 1873, upon his return from a meteorology conference in Vienna, yet another victim of the fourth nineteenth-century cholera pandemic (of 1863-1875). He was only in his mid-40s.

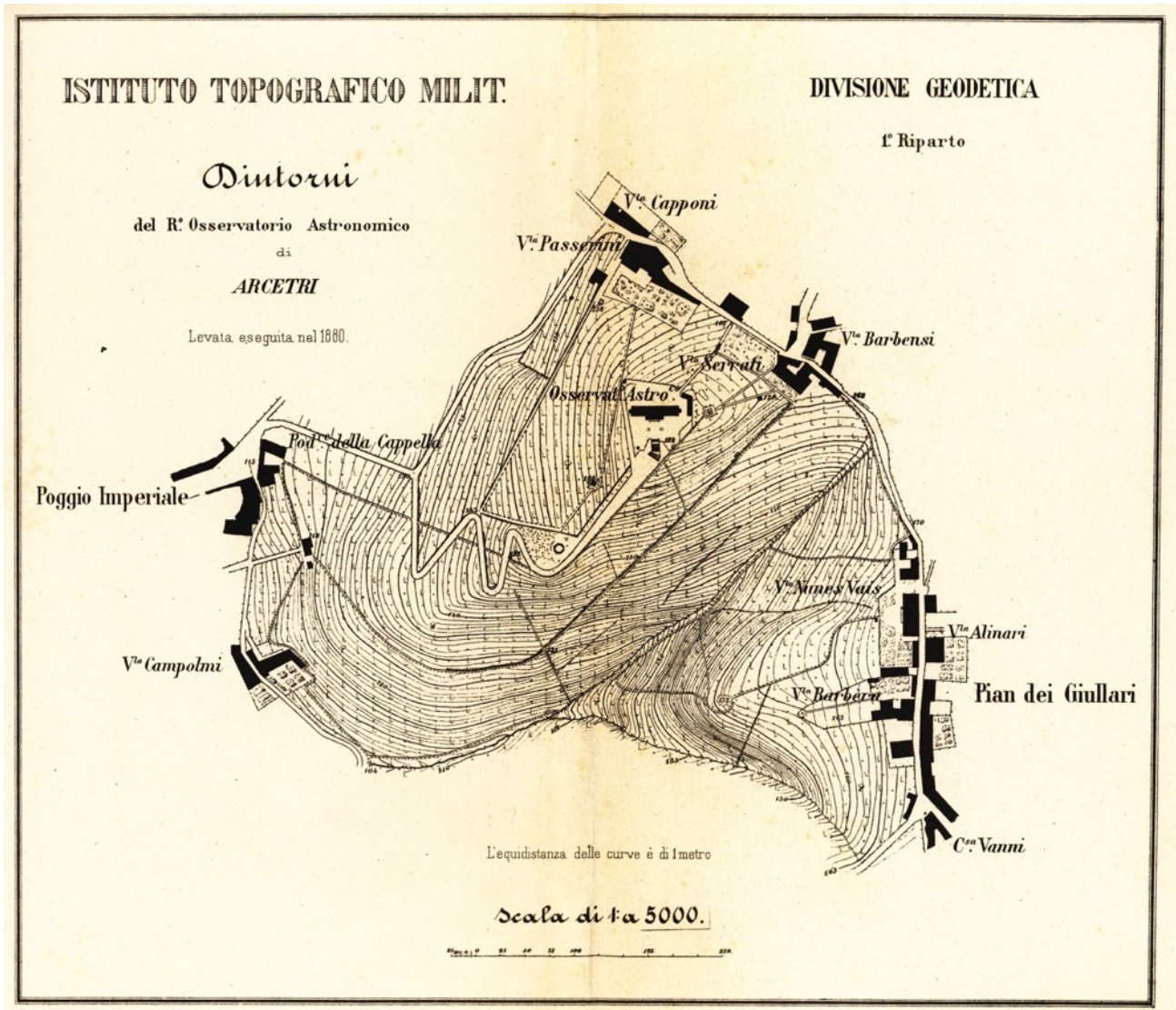

**Figure 18**. Arcetri Observatory and surrounds in 1880 (Archivio Storico INAF-Osservatorio Astrofisico di Arcetri). The circular wall of the 1869 pavilion still existed, but was never put into use for the magnetic instruments (Bianchi et al. 2011).

## 9 EPILOGUE

With the death of Donati, the first phase of the life of the Arcetri Astronomical Observatory can be considered as concluded. At the end of 1873 the authorities of the Institute almost succeeded in appointing the eminent astronomer Giovanni Virginio Schiaparelli as Director, but in January 1874 he declined the offer because of family reasons. It was then decided to leave the Directorship vacant, in order to save money and complete the instrumentation. Although this was intended as a temporary solution, the Observatory remained without a Director for 20 years. In the meanwhile, the only astronomer working at Arcetri was Tempel, who came from Brera Observatory at Schiaparelli's suggestion and employed at the end of 1874 (Bianchi et al. 2011).

Tempel had to work with incomplete instrumentation and in a building that was rapidly deteriorating. Rainwater leaks became an increasing problem, and the roof of the meridian hall had to be reinforced with posts and the West terrace covered with tiles. Other damage followed in the east wing. Most of the problems were due to lack of maintenance, which had ceased when there was a legal dispute between the Ministry, and the contractor Berti and the architect Falcini. Both Berti and Falcini wanted to be paid, but the Ministry attributed all the construction faults to them. After a solution of the dispute was reached, and further delays, a major refurbishment of the Observatory started in 1889 and a new Director, Antonio Abetti (1846-1928), was appointed at the end of 1893.

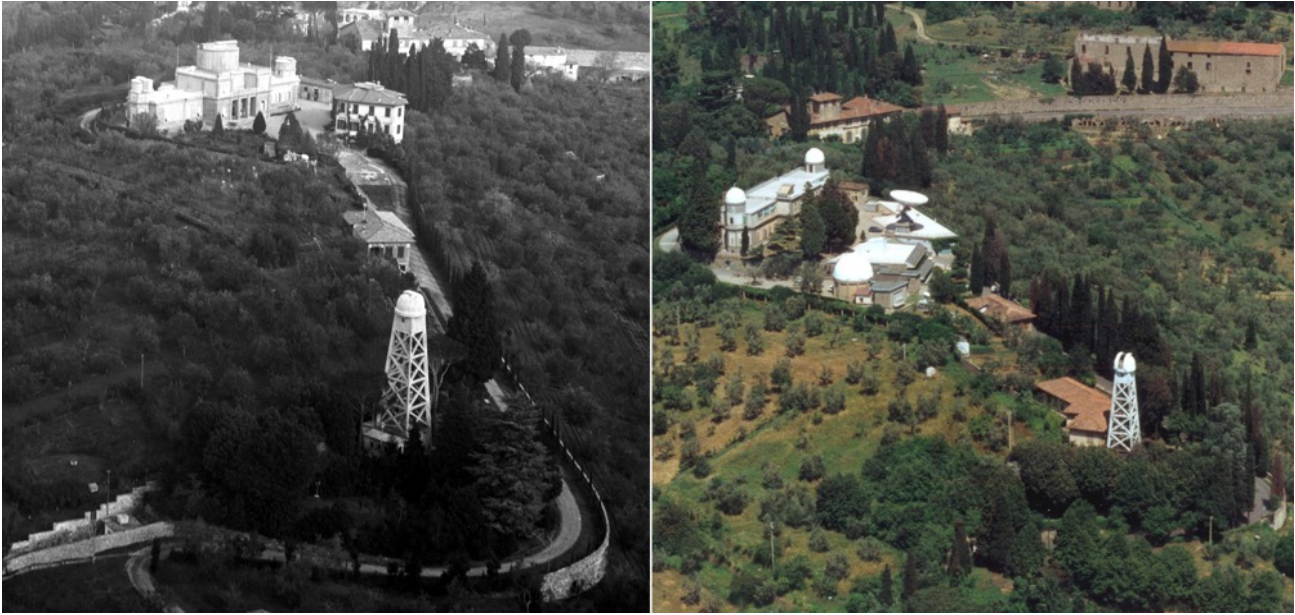

**Figure 19:** Aerial views of Arcetri Observatory in 1933 (left) and 1985 (right). The Solar Tower, in front, stands where Donati had the first installation of the Amici telescope (Figure 14). In a series of works started in 1959, the dome was moved to an external pavilion and the East and West wings of the main building were raised by one floor (Archivio Storico INAF-Osservatorio Astrofisico di Arcetri).

Abetti modified the mounting of the Amici telescope, which finally had a clock drive, graduated circles, and a metal tube, and he substituted the wooden cylindrical dome with one covered by metal. He obtained a Bamberg transit instrument and dreamed of buying the large meridian circle that Florence never had. However, in 1921 Arcetri became an "Astrophysical Observatory" under the Directorship of his son Giorgio (1882-1982) and the funds saved for the circle were used for the construction of the Solar Tower, which was inaugurated in 1925 (Bianchi et al. 2011). That same year, the Amici telescope was equipped with a new 36-cm (14-inch) Zeiss objective. The 1840s Amici doublet is now part of the historic collection of the Observatory, while the original mahogany tube is on display in the Museum Galileo in Florence. The telescope is still in use today for public observations in a dome detached from the main building (Figure 19).

A standard bearer of astrophysics in Italy (Bianchi 2021), Arcetri was dedicated for most of the 20[th] century to solar physics, while in the last four decades opening itself to several other research and technological fields. With over a hundred people working every day on the hill (including researchers, technicians, postdocs, students and administrative staff), the Arcetri Astrophysical Observatory continues to thrive as it awaits the celebration of the 150[th] Anniversary of its second inauguration, in 2022.

**10 NOTES**

1.	All English-language translations were made by the author.
2.	The current monetary values were estimated using the conversion tables provided by the Italian *Istituto Nazionale di Statistica* (https://www.istat.it/it/archivio/243273).
3.	The 14[th] century Franciscan Basilica of Santa Croce (Holy Cross) in Florence had been the burial place of distinguished citizens: among them Niccolò Machiavelli (1469-1527), Michelangelo Buonarroti (1475-1574) and Galileo Galilei (1574-1642). Starting with the remains of Piedmontese poet, but Florence resident, Vittorio Alfieri (1749-1803), from the beginning of the 19[th] century the church received monumental burials and cenotaphs of several others notable Italians; it became known as the "temple of Italian glories" (in the words of the poet Ugo Foscolo [1878-1827], whose remains are also in Santa Croce). When contemplating the tombs in 1817, the French author Marie-Henri Beyle (1783-1842) was overcome with awe and experienced the condition which is now known, after his pseudonym, as Stendhal's syndrome (Berti 1993).

## 11 ACKNOWLEDGEMENTS

The author wishes to thank Alberto Meschiari, who kindly provided transcripts of the yet unpublished correspondence between Antinori and Amici; Andreas Verdun, for his help in identifying the National Observatory of Switzerland quoted by Matteucci with the Zurich Federal Observatory; Adriano Ciabani, for allowing the use of the photograph of L. G. de Cambray Digny from his private collection; and Antonella Gasperini and Daniele Galli, for sharing with the author the enthusiasm of a decade-long search for -seldom published- documentation.